\documentclass[journal=jacsat,manuscript=article]{achemso}

\usepackage[version=3]{mhchem} %

\usepackage[utf8]{inputenc} %
\usepackage[T1]{fontenc}    %
\usepackage{hyperref}       %
\usepackage{url}            %
\usepackage{booktabs}       %
\usepackage{amsfonts}       %

\usepackage{nicefrac}       %
\usepackage{microtype}      %
\usepackage{lipsum}
\usepackage{fancyhdr}       %
\usepackage{graphicx}       %
\usepackage{subcaption}
\usepackage[font=small,labelfont=bf]{caption}       %
\graphicspath{{figures/}}     %
\usepackage[table,xcdraw]{xcolor}
\usepackage{array}
\usepackage{lscape}
\usepackage{multirow}
\usepackage{colortbl}
\usepackage{colortbl}
\usepackage{placeins}
\usepackage{soul,color}
\usepackage{amssymb}
\usepackage{amsmath}
\usepackage{pifont}
\newcommand{\cmark}{\checkmark}
\newcommand{\xmark}{\ding{55}}
\usepackage[table,xcdraw]{xcolor}
\usepackage[normalem]{ulem}
\useunder{\uline}{\ul}{}
\usepackage{longtable}

\author{Sadman Sadeed Omee}
\affiliation[University of South Carolina]
{Department of Computer Science and Engineering, University of South Carolina, Columbia, SC, 29201}
\author{Lai Wei}
\affiliation[University of South Carolina]
{Department of Computer Science and Engineering, University of South Carolina, Columbia, SC, 29201}
\author{Sourin Dey}
\affiliation[University of South Carolina]
{Department of Computer Science and Engineering, University of South Carolina, Columbia, SC, 29201}
\author{Jianjun Hu}
\affiliation[University of South Carolina]
{Department of Computer Science and Engineering, University of South Carolina, Columbia, SC, 29201}
\email{jianjunh@cse.sc.edu}

\title[ParetoCSP2]{Polymorphism Crystal Structure Prediction with Adaptive Space Group Diversity Control}

\abbreviations{IR,NMR,UV}
\begin{document}

\singlespacing %
\begin{abstract}
Crystalline materials can form different structural arrangements (i.e. polymorphs) with the same chemical composition, exhibiting distinct physical properties depending on how they were synthesized or the conditions under which they operate. For example, carbon can exist as graphite (soft, conductive) or diamond (hard, insulating). Computational methods that can predict these polymorphs are vital in materials science, which help understand stability relationships, guide synthesis efforts, and discover new materials with desired properties without extensive trial-and-error experimentation.
However, effective crystal structure prediction (CSP) algorithms for inorganic polymorph structures remain limited. We propose ParetoCSP2, a multi-objective genetic algorithm for polymorphism CSP that incorporates an adaptive space group diversity control technique, preventing over-representation of any single space group in the population guided by a neural network interatomic potential. Using an improved population initialization method and performing iterative structure relaxation, ParetoCSP2 not only alleviates premature convergence but also achieves improved convergence speed. Our results show that ParetoCSP2 achieves excellent performance in polymorphism prediction, including a nearly perfect space group and structural similarity accuracy for formulas with two polymorphs but with the same number of unit cell atoms. Evaluated on a benchmark dataset, it outperforms baseline algorithms by factors of 2.46-8.62 for these accuracies and improves by 44.8\%-87.04\% across key performance metrics for regular CSP. Our source code is freely available at \url{https://github.com/usccolumbia/ParetoCSP2}.
\end{abstract}

\label{sec:intro}

Polymorphism, the ability of a material to exist in multiple crystal structures with identical chemical composition but different atomic arrangements, represents a fundamental phenomenon in materials science with far-reaching implications \cite{buerger1937crystal,brog2013polymorphism,cruz2015facts}. This phenomenon is particularly significant for inorganic materials, where polymorphic diversity can dramatically influence functional properties including electronic structure, optical behavior, mechanical strength, and chemical reactivity \cite{gentili2019polymorphism}. Understanding and predicting polymorphism is crucial not only for advancing our theoretical knowledge of crystalline materials but also for practical applications spanning catalysis, energy storage, electronics, and pharmaceuticals.

The technological significance of polymorphic materials stems from the ability to access diverse material properties without changing the chemical composition. For example, silicon dioxide (SiO$_2$) exhibits distinct polymorphs: Quartz, Cristobalite, and Tridymite, each with different thermal stability and optical properties, making them essential for applications in electronics, optics, ceramics, and high-temperature materials~\cite{a-quartz1,a-quartz2,a-quartz3,cristo-1,cristo-2}. Similarly, carbon polymorphs ranging from graphite to diamond show a remarkable variation in hardness, thermal conductivity, and electronic character despite identical elemental composition~\cite{haggerty1999diamond,li2009superhard}. 
In the energy storage domain, various polymorphs of electrode materials, such as Li$_4$Ti$_5$O$_{12}$ present different lithium-ion diffusion pathways, directly impacting battery performance~\cite{huang2019li}. These examples underscore how polymorphic diversity enables material property engineering in multiple technological fields. Accurate prediction of polymorphic crystal structures is essential for optimizing material properties and alleviating risks associated with undesired polymorphic transitions, which can compromise structural stability and performance. For this, it is a critical consideration in fields such as pharmaceuticals, materials science, and semiconductors~\cite{nikhar2022reliable}. The challenges associated with polymorphism arise from the need to predict and distinguish between closely related structures, often separated by small lattice energy differences of approximately 10-15 kJ/mol, which require highly accurate computational methods~\cite{price2008computational,gavezzotti2007molecular}.

\begin{figure}[!htb]
    \centering
    \includegraphics[width=1\linewidth]{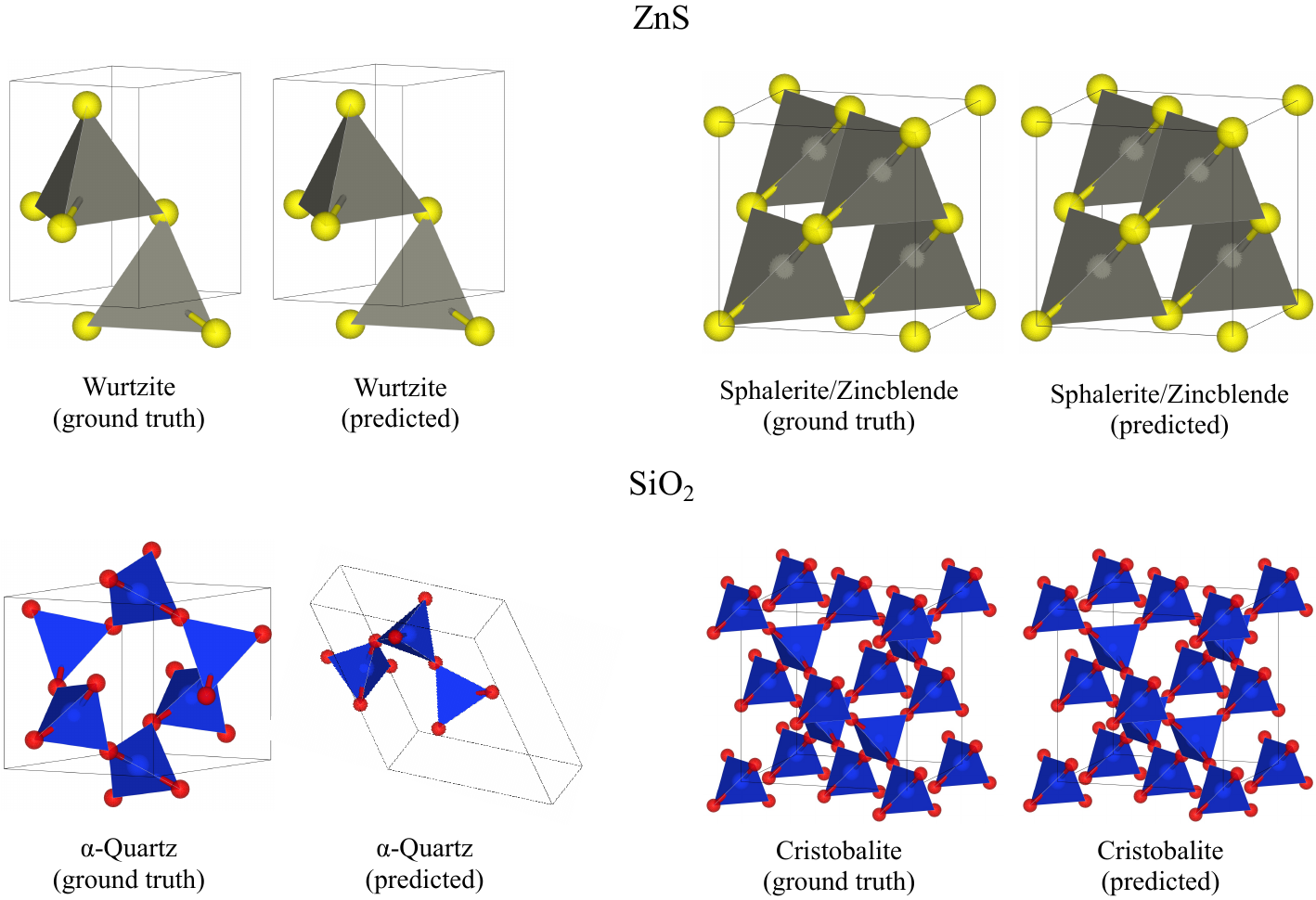}
    \caption{\textbf{Examples of polymorphic structure prediction by our algorithm.} Given a chemical formula, ParetoCSP2 aims to recover multiple known polymorphs. Both ground truth polymorphic structures and predicted polymorphic structures for ZnS and SiO$_2$ are shown. For ZnS, both Wurtzite and Sphalerite (Zincblende) polymorphs were successfully predicted. For SiO$_2$, our algorithm correctly identified Cristobalite but failed to recover the $\alpha$-Quartz structure, demonstrating both the capabilities and current limitations of the approach in capturing complete polymorphic diversity.}
    \label{fig:general_poly}
\end{figure}

Inspired by these potential applications, here we aim to develop an effective polymorphism crystal structure prediction algorithm for polymorph material discovery. 
Fig.~\ref{fig:general_poly} shows some examples of popular and widely used polymorphs in different industries, and some polymorph predictions by our algorithm (see Subsection ``Performance evaluation for polymorphism prediction''). Although crystal structure prediction (CSP) is a promising approach for material discovery \cite{oganov2010evolutionary,oganov2019structure}, accurate prediction of polymorphic crystal structures remains a difficult challenge despite significant advances in computational materials science.

First, CSP involves conducting a combinatorial search through a vast configuration space to identify the most thermodynamically stable structure, where the number of possible configurations grows exponentially with the number of atoms in the unit cell of the crystal~\cite{oganov2006crystal}. Second, this complexity is compounded by the computational cost of evaluating energy landscapes, often requiring first-principles calculations~\cite{uspex,airss,woodley2008crystal,calypso}. Consequently, the search for the global minimum on the potential energy surface (PES) is formally classified as an NP-hard computational problem~\cite{airss}. Traditional methods, such as X-ray diffraction (XRD)~\cite{bragg1913reflection,bragg1913structure}, while accurate, are immensely time-consuming, costly, and often impractical for difficult-to-synthesize materials.

Several notable CSP algorithms have been developed for crystal structure prediction, such as USPEX~\cite{uspex}, CALYPSO~\cite{uspex}, and AIRSS ~\cite{airss}, each implementing different search strategies to explore the vast configuration space. All these algorithms' dependence on expensive \textit{ab initio} calculations makes them infeasible for high-throughput materials discovery tasks. Recently, deep learning methods~\cite{megnet,omee2022scalable,m3gnet,chgnet} have been used to replace density functional theory (DFT) calculations to speed up structure prediction processes, as done in modern CSP algorithms such as GN-OA ~\cite{gnoa} and ParetoCSP ~\cite{paretocsp}. The former employs a graph neural network (GNN)~\cite{megnet} to predict formation energies and combines it with an optimization algorithm, such as Bayesian optimization (BO) and particle swarm optimization (PSO) to guide structure search, offering improved speed, but often struggling with high rates of invalid structure generation and low accuracy. ParetoCSP uses a multi-objective genetic algorithm (GA) NSGA-III~\cite{nsga3-1} enhanced by the age-fitness pareto optimization~\cite{afpo} technique to achieve higher search capability. However, it suffers from its structural convergence to structures with very few space groups. AlphaCrystal, devised by Hu et al.~\cite{alphacrystal} is a deep learning-based method that predicts crystal contact maps and reconstructs structures using genetic algorithms. However, its performance depends heavily on accurate predictions of space group and lattice parameters.

Traditional CSP algorithms often struggle to sample the vast configurational space efficiently while maintaining sufficient diversity in the predicted structures, particularly with respect to space group representation. Many algorithms tend to converge toward the most thermodynamically stable structures, overlooking metastable polymorphs that may offer superior functional properties for specific applications. Furthermore, the energy differences between polymorphs are often within the margin of error for density functional theory calculations, complicating the reliable ranking of candidate structures. These challenges highlight the need for more sophisticated approaches that can efficiently explore the complex energy landscape while preserving the diversity in the predicted polymorphic structures.

Despite the widespread use of current CSP algorithms, most of them do not take polymorphism into account to find energetically sub-optimal stable structures, but labeling them as incorrect predictions. The only exception is an improved version of CALYPSO \cite{spai}, which uses divide-and-conquer heuristic rules to control the diversity of the structure population in terms of space groups during its PSO search.  
In fact, most of the existing work on polymorphism crystals is directed at molecular crystals~\cite{price2008computational,bernstein2020polymorphism}. Significant progress toward this goal has been effectively demonstrated through the international blind tests on crystal structure prediction conducted by the Cambridge Crystallographic Data Centre (CCDC)~\cite{ccdc-2,ccdc-4,ccdc-7}. Several other approaches have been proposed to predict low-energy polymorphs of individual compounds, including Polymorph Predictor~\cite{verwer1998computer}, PROMET~\cite{promet}, MOLPAK~\cite{molpak}, MPA~\cite{mpa}, etc.~\cite{cruz2015facts,desiraju1997crystal,price2004computational}. Several machine learning (ML)-based approaches for polymorphism prediction for organic molecules have also been proposed~\cite{oliynyk2017disentangling,mcdonagh2019machine,collins2010comparing}. However, these methods are mainly focused on predicting the polymorphs of a single molecular compound or a single group of compounds, and a combination of these search methods has often been used to predict the polymorphs.

In this article, we propose ParetoCSP2, a multi-objective genetic algorithm (GA) to address one of the biggest challenges in polymorphism CSP, which is ensuring the diversity of space groups in the population or final candidates~\cite{spai}. To address this limitation, we incorporate the frequency of each space group in the GA population as a new optimization criterion in our multi-objective genetic algorithm, alongside energy and genotypic age~\cite{afpo}. By minimizing the maximum number of individuals from each space group, our approach tends to evolve diverse structures with lower energy and age within the Pareto front. Additionally, we replace the previous randomly generated initial crystals with a robust structure generation method from the \texttt{PyXtal}~\cite{pyxtal} library. This modification increases the proportion of valid structures available for the genetic algorithm, thereby enhancing the probability of discovering optimal crystal configurations. Furthermore, we implement structure relaxation after each generation to provide better guidance during the search process and achieve faster convergence.

\section{Methods}
\label{sec:method}

\subsection{ParetoCSP2: a polymorphism CSP algorithm}

Our ParetoCSP2 is developed based on our previously developed ParetoCSP~\cite{paretocsp} algorithm for generic crystal structure prediction. We keep the notation the same as that in the ParetoCSP paper. The input of the CSP problem is the elemental composition, denoted by $\{c_i\}$ where $i=1,2,...,n$ which denotes the index of an atom in the unit cell and $c_i$ denotes the $i$-th atom's element. The lattice parameters ($L)$ are composed of six lattice constants: a, b, c representing the lengths in the unit cell and $\alpha$, $\beta$, $\gamma$ representing the angles in the unit cell. The space group representing the crystal symmetry is denoted by $S$. The Wyckoff positions and the atomic coordinates are denoted by $\{W_i\}$ and $\{R_i\}$, respectively, where $W_i$ and $R_i$ denote the Wyckoff position and the fractional coordinate of the atom indexed by $i$ where $i=1,2,...,n$, respectively. The periodic nature of a crystal can be represented by the space group, lattice parameters, Wyckoff positions, and atomic coordinates. A flowchart of the algorithm with an example crystal CeCr$_2$Si$_2$C is shown in Fig.~\ref{fig:flowchart} and the pseudocode of the algorithm is provided in Supplementary Note S1.1.

\begin{figure}[!htb]
    \centering
    \includegraphics[width=1\linewidth]{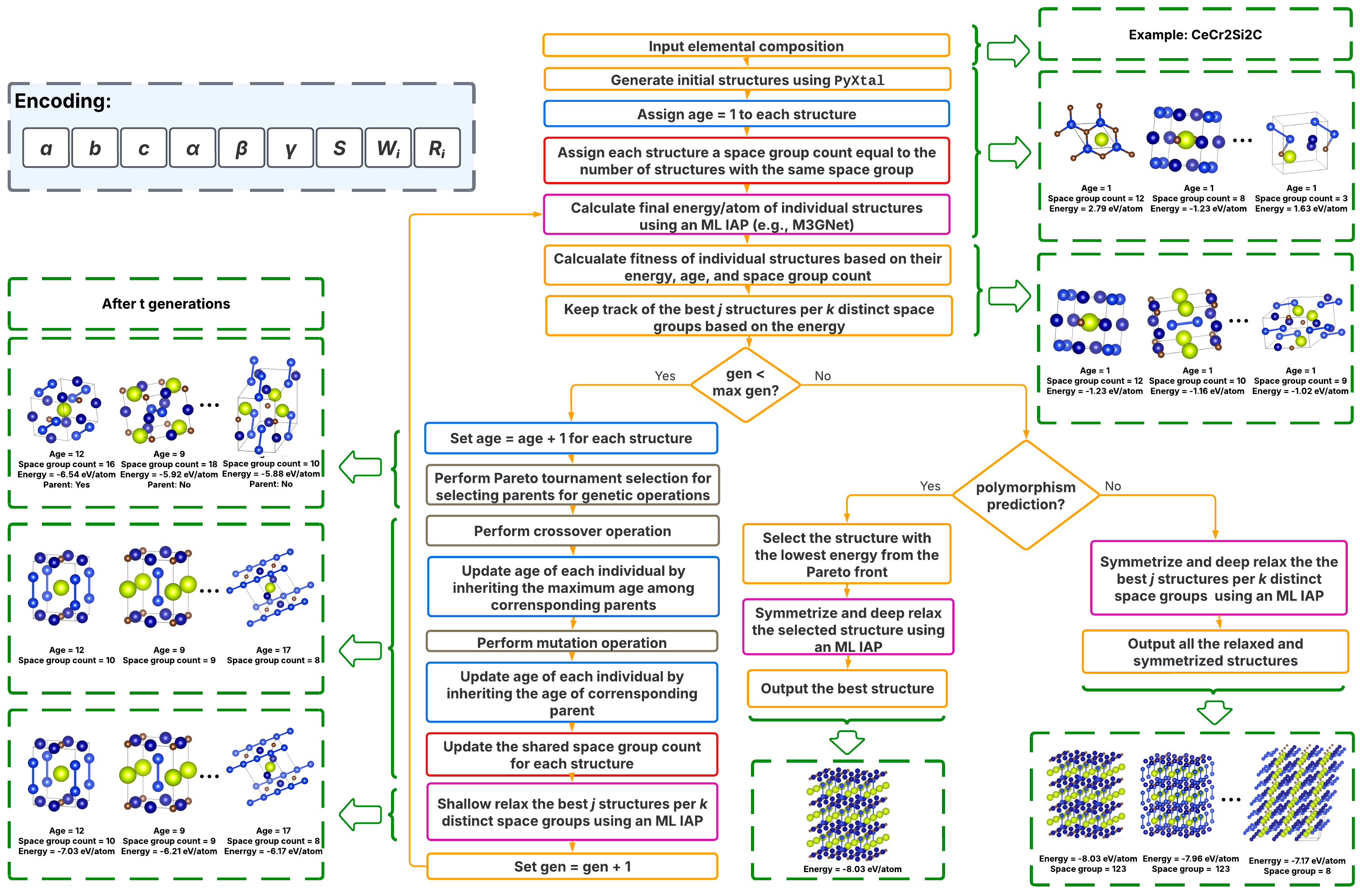}
    \caption{\textbf{Flowchart of the ParetoCSP2 algorithm.} Starting from a given chemical composition (e.g., CeCr$_2$Si$_2$C), the algorithm begins by generating an initial population of structures using \texttt{PyXtal}, assigning a genotypic age of 1 to them, and computing their shared space group count based on the number of structures sharing the same space group within the population. Matching colors are used to indicate functionally similar steps for better visual understanding. Each generation proceeds through the following steps: energy prediction, fitness evaluation based on three objectives (energy, genotypic age, and shared space group count), parent selection, genetic operations (crossover and mutation), age and shared space group count updates, and shallow relaxation of all structures. ParetoCSP2 tracks the best $j$ structures per $k$ distinct space groups at each generation. After reaching a maximum of $\mathcal{G}$ generations, the structure with the lowest predicted energy from the multi-objective Pareto front is selected, further relaxed, and symmetrized to output the final predicted structure. For polymorphism prediction, all tracked structures are also relaxed and symmetrized to produce a set of potential polymorphs. The top-left part of the flowchart shows the genetic encoding scheme, which includes lattice parameters ($a$, $b$, $c$, $\alpha$, $\beta$, $\gamma$), the space group ($S$), Wyckoff position combinations ($W_i$), and atomic coordinates ($R_i$) for atoms indexed by $i$.}
    \label{fig:flowchart}
\end{figure}

For a given $\{c_i\}$, the number of possible crystal structures for different combinations of $S, L, \{W_i\},$ and $\{R_i\}$ can be astronomical. So, a pragmatic solution is to iteratively sample candidate structures assuming that the optimal solution will be among them. For this purpose, we used a genetic algorithm (GA) named NSGA-III enhanced by the age-fitness Pareto optimization (AFPO) method in our version - ParetoCSP. In this version, we first incorporate the maximum space group count of each possible space group as an independent optimization criterion. This ensures diversity of space groups in the population by not allowing too many structures with the same space group. For example, for SrTiO$_3$, ParetoCSP2 identifies 67 distinct space group structures after 20 generations for a population size of 100, compared to only 15 in ParetoCSP. Moreover, we employ an ML-based interatomic potential (IAP) to predict the structure energy to guide the GA search towards the global optima.

The process begins with the generation of an initial set of $n$ structures. Instead of relying on random generation, as in the previous version, which often produces invalid crystals, i.e., structures with high energy, an efficient method from the PyXtal library is employed. This method takes the composition and the space group as inputs and produces candidate structures based on the specified space group. PyXtal generates more valid crystal structures at the outset, increasing the likelihood of identifying the optimal crystal structure in later iterations. Moreover, PyXtal enables the algorithm to generate diverse space group structures for the initial population, boosting enhanced space group exploration from the very beginning. Subsequently, we encoded them in vectors for genetic operations and assigned a genotypic age of 1 to the new structures. After that, we calculate the occurrence of space group for each space group from $P1$ to $P230$ and assign to each individual the shared space group count of its own space group. 

In the implementation of ParetoCSP, the correct space group often appeared only after relaxation of the final candidate structures, which prevented the algorithm from getting better guidance during the search. To address this issue, we use M3GNet~\cite{m3gnet} or CHGNet~\cite{chgnet} to relax all candidate structures after each generation in ParetoCSP2. Specifically, we decode the encoded vectors representing all individuals into Pymatgen structures, relax them using M3GNet or CHGNet, and then re-encode them back into fixed-size vectors. Relaxing all structures accelerates the search for optimal crystals, which is frequently identified within the first 1–10 generations. This process is memory-intensive, and we handle this by implementing efficient programming techniques such as memory-efficient data structures and garbage collection methods. We then use an ML IAP (M3GNet or CHGNet) to predict their final energy/atom. The algorithm then evaluates the fitness of each individual by considering all three factors: energy, genotypic age, and the diversity of space groups represented. Finally, the algorithm verifies whether the termination condition has been met.

The default terminating condition for ParetoCSP2 is a fixed threshold of generations $\mathcal{G}$. If the algorithm has not met the terminating condition, it increments the age of all individuals by 1. Then the algorithm performs the genetic operations: crossover and mutation. The Pareto tournament selection method (default tournament size: 2) is utilized to select parent individuals for subsequent genetic operations. After performing the genetic operations, we update the age of each individual by the following scheme: after crossover, an individual inherits the maximum age of its parents, while after mutation, an individual directly inherits the age of its parent. The incorporation of age as an independent optimization criterion promotes population diversity by maintaining a balance of younger and older individuals, thus reducing the likelihood of premature convergence to local optima. Analogous to the greater survival potential of a young lion compared to an older one in the wild, younger solutions are prioritized over those with higher ages, ensuring a dynamic and exploratory search process. After that, we update the shared space group count for each individual and relax all the structures. Finally, we predict the final energy/atom for each structure using either M3GNet or CHGNet. This process results in $n$ new individuals for subsequent generations, after which the algorithm calculates the fitness of all individuals.

The entire process is iteratively repeated until the termination condition is satisfied, resulting in a set of $\mathcal{F}$ non-dominated solutions on the Pareto front. We further relax these solutions using the ML IAP, often yielding more precise structures. Finally, we select the structure with the lowest energy and apply a symmetrization operation and output it as the final structure. Additionally, for polymorphism prediction, ParetoCSP2 the algorithm tracks $j$ structures for each $k$ distinct space groups ($k$ lowest energy structures, each with a different space group), saving a total of $jk$ structures after each generation. In our experiments, we set $j = 3$ and $k = 10$. Tracking structures across $k$ different space groups ensures that the algorithm can predict energetically stable structures with the same composition but different structural formations. Additionally, since crystal polymorphism may occasionally occur within the same space group, the algorithm retains the $j$ best structures for each of the $k$ space groups to account for such rare occurrences. Studies have shown that multiple space group structures tracked during the search process eventually converge to the same structure after relaxation for most algorithms~\cite{spai,hessmann2025accelerating}. But our relaxation of all structures after each generation automatically handles that issue.

\subsection{Age-fitness Pareto optimization (AFPO)}
To maintain structural diversity and prevent premature convergence in the genetic search process, ParetoCSP2 uses the AFPO algorithm by Schmidt and Lipson~\cite{afpo}. The \textit{age} of an individual refers to the number of generations since the oldest part of its genetic material appeared for the first time in the population. In a general multi-objective optimization setting, each candidate solution \( x_i \) in the population is associated with a fitness vector $f(x_i) = (f_1(x_i), f_2(x_i), \ldots, f_k(x_i), a(x_i))$, where \( f_j(x_i) \) for \( j = 1, 2, \ldots, k \) represents the different optimization objectives (e.g. cost, energy, violation of restrictions) and \( a(x_i) \in \mathbb{N} \) denotes the genotypic age of the individual. A solution is considered Pareto-optimal if it achieves higher fitness and lower age compared to other solutions.

At each generation, all individuals increment their age by one, and any newly generated offspring inherit the maximum age of their parents, i.e., \( a_{\text{child}} = \max(a_{\text{parent}_1}, a_{\text{parent}_2}) \). A tournament selection of size \( K \) is used, where \( K \) individuals are sampled, and dominated ones are discarded based on Pareto front ranking in the two-dimensional objective space. Genetic variation operators such as crossover and mutation are applied to the surviving parents to generate offspring. The offspring replace older individuals based on fitness and age, while a small number of random individuals with age \( a = 1 \) are introduced each generation to encourage population diversity. This multi-objective optimization strategy enables ParetoCSP2 to perform an effective global search over the crystal configuration space.

\subsection{NSGA-III: multi-objective GA}
NSGA-III~\cite{nsga3-1} is a multi-objective evolutionary algorithm designed to handle optimization problems involving three or more objectives. Let the parent population in generation $i$ be denoted by $P_i$, and the offspring population generated by genetic operations be $Q_i$. The combined population is defined as $R_i = P_i \cup Q_i$, which is then sorted into a series of non-dominated fronts $F_1, F_2, \ldots$. The new population $P_{i+1}$ is formed by sequentially adding fronts $F_1$ through $F_k$ such that $\sum_{j=1}^{k} |F_j| \leq N$, where $N$ is the population size. If including $F_{k+1}$ exceeds the population limit, only a subset of individuals from $F_{k+1}$ is selected to fill the remaining slots.

To preserve population diversity, NSGA-III introduces a set of predefined reference points $\mathcal{R} = \{ r_1, r_2, \ldots, r_m \} \subset \mathbb{R}^M$, where $M$ is the number of objectives. Each solution $x \in F_{k+1}$ is normalized and associated with the reference point $r_j \in \mathcal{R}$ that minimizes the perpendicular distance from $x$ to the reference line defined by the ideal point and $r_j$. The selection of $F_{k+1}$ is then performed using a niching procedure, which prioritizes solutions associated with sparsely represented reference points in $P_{i+1}$. This reference-point-guided niching strategy allows NSGA-III to maintain a well-distributed set of solutions along the Pareto front, especially in high-dimensional objective spaces. For further algorithmic details, the readers are referred to~\cite{nsga3-1}.

\subsection{ML inter-atomic potential (ML IAPs)}
\subsubsection{M3GNet}
M3GNet~\cite{m3gnet} is a GNN-based IAP designed for accurate and efficient energy prediction and structural relaxation in crystal structure prediction (CSP). Unlike earlier models such as MEGNet~\cite{megnet}, M3GNet explicitly incorporates 3-body interactions and utilizes atomic coordinates and the $3 \times 3$ lattice matrix, enabling the computation of tensorial quantities like forces and stresses via auto-differentiation. It extends traditional GNN architectures by including bond angle information, a many-body computation module, and gated multilayer perceptrons (MLPs) for atomic energy prediction. Trained on a large dataset containing both stable and unstable structures, M3GNet is particularly effective in capturing the energetic landscape of intermediate crystal configurations. Its enhanced capabilities make it well-suited for scalable, high-throughput CSP tasks.

\subsubsection{CHGNet}
CHGNet(Crystal Hamiltonian Graph Network)~\cite{chgnet} is a pretrained universal ML IAP designed for charge-informed atomistic modeling, utilizing a GNN to predict energies, forces, stresses, and magnetic moments. Unlike M3GNet, which focuses on general interatomic potentials for the periodic table, CHGNet explicitly incorporates electronic degrees of freedom, particularly spin-polarized magnetic moments (magmoms), allowing accurate modeling of complex electron interactions and charge dynamics in materials like $\mathrm{Li}_x \mathrm{MnO}_2$ and $\mathrm{Li}_x \mathrm{FePO}_4$. CHGNet is trained on the Materials Project Trajectory (MPtrj) Dataset, comprising over 1.5 million inorganic structures from density functional theory calculations, covering diverse chemistries excluding only the noble gases and actinoids. This makes CHGNet a powerful and general-purpose surrogate model for accelerating high-throughput materials discovery, especially in chemically and electronically complex systems.

\subsection{Evaluation metrics}
While numerous structural evaluation metrics have been developed for molecular structure prediction~\cite{mol1,mol2,mol3}, the field of crystal structure prediction (CSP) has historically lacked rigorous quantitative evaluation tools. Many prior studies depended mainly on structural comparison by manual inspection and \textit{ab initio} formation energy analysis to evaluate performance~\cite{uspex,calypso,airss}. However, these methods do not quantitatively measure the similarity between the predicted and ground truth structures, and they also fail to effectively compare predictions that closely approximate the correct solution across different algorithms. To address these limitations, Wei et al.~\cite{cspmetric} introduced a set of quantitative metrics specifically designed for the CSP assessment of various algorithms. In this work, we adopt five of these metrics to benchmark the performance of our proposed ParetoCSP2 and other algorithms. They are briefly listed in Table~\ref{tab:metrics} and described in detail in Supplementary Note S1.2:

\begin{table}[!htb]
\centering
\caption{\textbf{Brief description of evaluation metrics used in this work.} Each metric captures a specific structural or energetic similarity between predicted and ground truth structures.}
\label{tab:metrics}
\begin{tabular}{l c c}
\toprule
\toprule
\textbf{Metric name} & \textbf{Brief description} & \textbf{Unit}\\
\midrule
\midrule
Energy Distance (ED) & Difference in structure energy & eV/atom\\\hline
Sinkhorn Distance (SD) & \multicolumn{1}{c}{\begin{tabular}[c]{@{}c@{}}{Structural similarity via}\\{optimal transport}\end{tabular}} & Angstroms (\r{A})\\\hline
Chamfer Distance (CD) & \multicolumn{1}{c}{\begin{tabular}[c]{@{}c@{}}{Average point-wise distance among}\\{predicted and ground-truth atom positions}\end{tabular}} & Angstroms (\r{A})\\\hline
Hausdorff Distance (HD) & \multicolumn{1}{c}{\begin{tabular}[c]{@{}c@{}}{Maximum deviation among}\\{predicted and ground-truth atoms}\end{tabular}} & Angstroms (\r{A})\\\hline
Fingerprint Distance (FP) &  \multicolumn{1}{c}{\begin{tabular}[c]{@{}c@{}}{A neural network-based~\cite{zimmermann2020local}}\\{feature vector similarity}\end{tabular}} & Unitless\\
\bottomrule
\bottomrule
\end{tabular}
\end{table}

Determining the appropriate thresholds for these metrics to qualify as an exact match is challenging, as the thresholds may vary depending on the crystal. However, these metrics can give us a comparative rank among different algorithms, with lower values indicating better structural predictions. Additionally, to properly address the exact match of an algorithm, we used two other metrics in conjunction - space group match, and \texttt{Pymatgen}'s StructureMatcher match. 

Space group matching verifies whether a predicted and a ground truth structure share the identical space group, as if the space groups do not match, we can rule them out for an exact match. Pymatgen's StructureMatcher determines whether two structures match by comparing their lattice parameters and atomic positions after reducing them to primitive cells, optionally scaling to the same volume, and aligning atoms to a common origin. It computes the root mean squared (RMS) displacement between atomic sites and declares a match if the normalized maximum RMS displacement is within a specified site tolerance. However, StructureMatcher does not consider space group match, and so a prediction might be considered a match in StructureMatcher, but not a space group match depending on the threshold values set. For this reason, we used both metrics in conjunction to determine exact matches of the predictions. We use a fractional length tolerance of 0.2, a site tolerance of 0.3, and an angle tolerance of 5 degrees for evaluating performances using StructureMatcher.

For evaluating polymorphism prediction results, we use coverage rates of space group match rate and StructureMatcher for each chosen formula. Given a chemical formula with $p$ different polymorphs and our algorithm successfully predicted the space group of $q$ polymorphs and had successful StructureMatcher matches with $r$ polymorphs, we define the space group coverage and the StructureMatcher coverage for this formula as $\frac{q}{p} \times 100\%$ and $\frac{r}{p} \times 100\%$, respectively.

\subsection{Benchmark datasets}
\subsubsection{Benchmark dataset for polymorphism CSP}

To evaluate the performance of our ParetoCSP2 algorithm in predicting polymorphic structures, we analyzed the space-group coverage rate and the StructureMatcher coverage rate across a diverse set of chemical compositions that have polymorphic forms. The definition of these coverage rates is described in the ``Evaluation metrics'' subsection. For this, we constructed a test set from the Materials Project database \cite{10.1063/1.4812323}. We only selected binary, ternary, or quarternary crystalline materials with $\leq 20$ atoms in their unit cell and having $\leq 10$ polymorphs with identical numbers of atoms. We want to mention here that the choice of using $\leq 10$ polymorphs was a matter of setting a hyperparameter value, not a limitation of our algorithm, as it is capable to predict any number of polymorphs. This gave us a set of crystals having polymorphic forms with the same number of atoms in their unit cell. In total, we found only two formulas with seven polymorphs, two formulas with eight polymorphs, and only one formula with 10 polymorphs. We chose a total of 50 formulas for evaluation that included all five of the mentioned higher-occurrence polymorph formulas. The details of the polymorphs chosen are presented in Supplementary Table S1.

\subsubsection{Benchmark dataset for regular CSP}
Here we aim to evaluate how ParetoCSP2's performance for regular CSP has been improved due to its better structure diversity control. Traditionally, CSP algorithms have been tested on small, narrowly defined datasets \cite{spai,uspex,calypso,gnoa,han2025efficient}, often focusing on only a few crystal structures or specific systems such as carbon allotropes or C-Si systems. Although these tests can show how an algorithm performs in a narrow setting, they do not provide enough insight into how well it would work across a wide variety of chemical compositions, symmetries, and structural patterns. Without a diverse benchmark test set, it is difficult to truly understand the reliability and usefulness of CSP algorithms for real-world material discovery.

To overcome the limitations of narrow benchmark sets, we adopt the recent work of Wei et al.~\cite{wei2024cspbench} (CSPBench), which comprises 180 inorganic crystal structures from the Materials Project database \cite{10.1063/1.4812323} spanning a wide range of chemical elements, space groups, and structural complexities. This benchmark test set was systematically designed to support fair and comprehensive comparisons among different CSP algorithms across a consistent and diverse evaluation set. The test set is divided into binary, ternary, and quarternary crystal groups that include varying difficulty levels to rigorously evaluate the performance of crystal structure prediction algorithms. For our study, we selected all structures from CSPBench that contain 20 or fewer atoms in their unit cells to ensure computational feasibility while maintaining broad coverage. We found a total number of 120 crystals that meet our requirements. Supplementary Table S2 shows the detailed information about the chosen 120 benchmark crystals, and the diversity of the elements used is shown in Supplementary Fig. S1. Supplementary Fig. S2 shows the distribution of crystal systems in the dataset and the corresponding formation energies by crystal systems.

\section{Results}
\subsection{Performance evaluation for polymorphism prediction}
In this work, we primarily focus on a specific case of polymorphism where polymorphs share identical numbers of atoms in the unit cell for a given chemical formula. However, our algorithm can be adapted for generalized polymorphism prediction for which we present several case studies. To evaluate our algorithm, we used coverage rates of the space group and StructureMatcher matches for each selected chemical formula. Space group coverage refers to the percentage of ground truth polymorphs for which the algorithm correctly predicted the space group, while StructureMatcher coverage indicates the percentage of ground truth polymorphs that are structurally similar to the predicted ones based on the Pymatgen's StructureMatcher method.

\begin{figure}[!htb]
    \centering
    \includegraphics[width=1\linewidth]{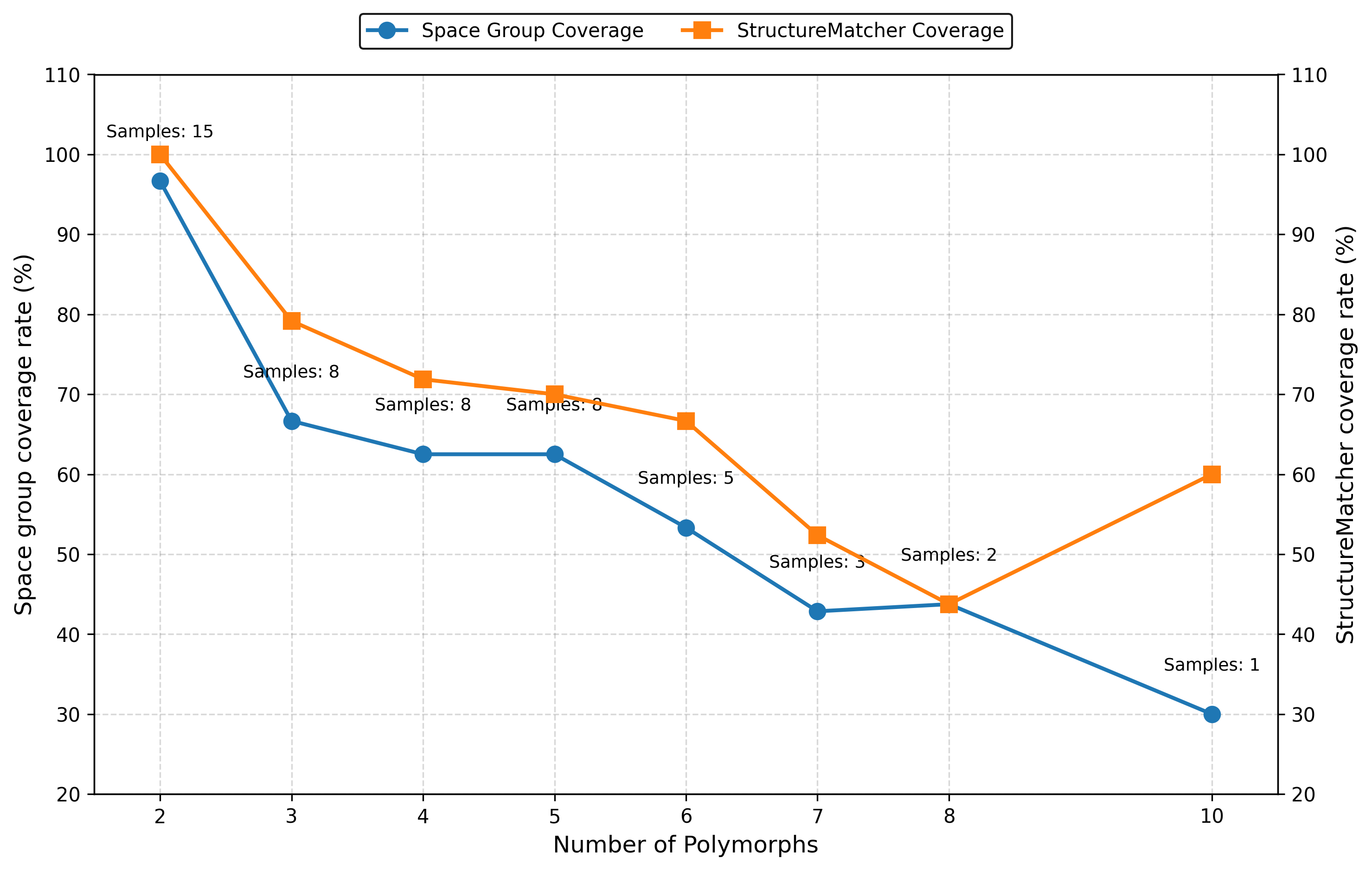}
    \caption{\textbf{Average polymorph coverage rates achieved by ParetoCSP2 across different levels of polymorphism complexity.} The x-axis denotes the number of known polymorphs for each compound class, while the y-axes represent the average coverage rates: space group match (blue, left axis) and Pymatgen StructureMatcher match (orange, right axis). Each point is annotated with the number of benchmark samples in that category. As the number of polymorphs increases, the average coverage rate tends to decline, indicating the increasing difficulty of capturing all distinct polymorphs. StructureMatcher generally provides slightly higher coverage due to its tolerance of structural variations, while space group match remains stricter.}
    \label{fig:avg_poly}
\end{figure}

The chemical formula-wise coverage rates for space group and StructureMatcher are presented in Table~\ref{tab:poly} and the average coverage rates for each polymorph class are shown in Fig.~\ref{fig:avg_poly}. They demonstrate that ParetoCSP2 successfully identified a significant portion of the polymorphic landscape, but its performance varied with the complexity of the polymorphic system. Some examples of successful predictions are shown in Supplementary Fig. S3. From the dataset, unit cell formulas that show only two polymorphs achieved near-perfect average space group coverage (96.67\%) and complete average StructureMatcher coverage (100\%). This indicates that ParetoCSP2 is highly effective in capturing polymorphism when there are only a limited number of structural variations.

However, as the number of polymorphs increased, the coverage rates decreased, reflecting the increasing complexity of predicting multiple energetically competitive crystal structures. For compositions with three polymorphs, the average space group coverage rate dropped to 66.67\%, while the average StructureMatcher coverage remained relatively high (79.17\%). The space group coverage rate remained the same (62.5\%) for chemical compositions with four and five polymorphs, and achieved 53.33\% average coverage rate for six polymorphs, while the average StructureMatcher coverage rates slowly declined from 71.88\%, and 70\% to 66.67\%, respectively. This suggests that while our method is capable of recovering polymorphs with distinct space groups, it underperformed with cases with an increasing number of polymorphs for a given unit cell formula. However, it is important to note that the accuracy followed a similar trend with respect to the number of available samples. Given more samples, the algorithm might achieve a better overall accuracy.

\renewcommand{\arraystretch}{1.4}
\begin{longtable}{l c c c}
\caption{\textbf{Polymorphism coverage rates of ParetoCSP2 for space group and StructureMatcher for the selected chemical formula.} The formulas represent the number of atoms present in the unit cell of the polymorphic structures. The formulas are divided by a horizontal line based on the number of polymorphs they can form. ParetoCSP2 achieved good prediction performance for formulas with 2-5 polymorphic forms. The overall performance decreased with increasing number of polymorphs, but this trend may be influenced by the limited number of samples available that met our selection criterion in higher polymorph classes.} \label{tab:poly} \\
\hline\hline
\noalign{\vskip 0.5mm}
\hline\hline
\multicolumn{1}{l}{\begin{tabular}[c]{@{}c@{}}\textbf{Unit cell}\\\textbf{formula}\end{tabular}} & \multicolumn{1}{c}{\begin{tabular}[c]{@{}c@{}}\textbf{Number of}\\\textbf{polymorphs}\end{tabular}} & \multicolumn{1}{c}{\begin{tabular}[c]{@{}c@{}}\textbf{Space group}\\\textbf{coverage rate (\%)}\end{tabular}} & \multicolumn{1}{c}{\begin{tabular}[c]{@{}c@{}}\textbf{StructureMatcher}\\\textbf{coverage rate (\%)}\end{tabular}} \\
\hline\hline
\noalign{\vskip 0.5mm}
\hline\hline
\endfirsthead

Ce$_1$Tl$_1$    & 2  & 100    & 100    \\
Er$_1$Ag$_1$S$_2$  & 2  & 50     & 100    \\
Er$_1$N$_1$     & 2  & 100    & 100    \\
Fe$_3$Sn$_1$    & 2  & 100    & 100    \\
Li$_1$Pd$_1$N$_1$  & 2  & 100    & 100    \\
Mg$_1$Ti$_1$    & 2  & 100    & 100    \\
Ni$_1$Hg$_1$    & 2  & 100    & 100    \\
Re$_1$Pt$_1$    & 2  & 100    & 100    \\
Tm$_1$Te$_1$    & 2  & 100    & 100    \\
V$_1$Fe$_1$     & 2  & 100    & 100    \\
Y$_1$As$_1$     & 2  & 100    & 100    \\
Si$_1$C$_1$     & 2  & 100    & 100    \\
Sr$_1$Te$_1$    & 2  & 100    & 100    \\
Y$_1$P$_1$     & 2  & 100    & 100    \\
Sn$_1$P$_1$    & 2  & 100    & 100    \\
\hline\hline
Ce$_1$Bi$_1$    & 3  & 100    & 100    \\
Mg$_1$O$_1$     & 3  & 100    & 100    \\
Ta$_1$Ru$_1$    & 3  & 100    & 100    \\
Al$_1$As$_1$    & 3  & 33.33  & 66.67  \\
Ba$_1$C$_2$     & 3  & 33.33  & 66.67  \\
Hf$_1$N$_1$     & 3  & 66.67  & 66.67  \\
Mn$_1$Sb$_1$    & 3  & 33.33  & 66.67  \\
Sm$_1$Hg$_2$    & 3  & 66.67  & 66.67  \\
\hline\hline
La$_3$Mg$_1$   & 4  & 50     & 100    \\
Mg$_1$Sn$_1$   & 4  & 75     & 100    \\
Ag$_2$O$_2$    & 4  & 25     & 25     \\
Cd$_1$Cl$_2$   & 4  & 75     & 75     \\
Ce$_1$Mg$_1$   & 4  & 75     & 75     \\
Mg$_2$Zn$_2$   & 4  & 75     & 75     \\
Os$_1$C$_1$    & 4  & 75     & 75     \\
Ti$_1$S$_2$    & 4  & 50     & 50     \\
\hline\hline
Ge$_1$Te$_1$   & 5  & 40     & 80     \\
Li$_1$Mg$_1$   & 5  & 80     & 80     \\
Mn$_1$O$_2$    & 5  & 80     & 80     \\
Cu$_3$N$_1$    & 5  & 60     & 60     \\
Mg$_1$Sb$_1$   & 5  & 60     & 60     \\
Mo$_1$N$_1$    & 5  & 60     & 80     \\
Rb$_1$S$_1$    & 5  & 60     & 60     \\
Mg$_3$Ga$_1$   & 5  & 60     & 60     \\
\hline\hline
Mg$_1$Ga$_3$   & 6  & 50     & 83.33  \\
Mg$_1$Al$_1$   & 6  & 66.67  & 66.67     \\
Mg$_1$Cd$_1$   & 6  & 50     & 50     \\
Mg$_1$Ga$_1$   & 6  & 50     & 66.67     \\
Ce$_3$Mg$_1$   & 6  & 50     & 66.67  \\
\hline\hline
Y$_1$Mg$_1$    & 7  & 42.86  & 85.71  \\
In$_1$Sb$_1$   & 7  & 42.86  & 42.86  \\
Ca$_3$Mg$_1$   & 7  & 42.86  & 28.57  \\
\hline\hline
Nb$_1$S$_2$    & 8  & 25     & 50     \\
V$_1$S$_2$     & 8  & 62.5     & 50     \\
\hline\hline
Na$_1$N$_3$    & 10 & 30     & 60     \\
\hline\hline
\noalign{\vskip 0.5mm}
\hline\hline
\end{longtable}

For compositions with six or more polymorphs, the trend becomes even more evident. The average space group coverage rate declined further to 42.86\% and 43.75\% for chemical formulas with seven and eight polymorphs, respectively, indicating that the algorithm experienced increasing difficulty in capturing the full range of structural diversity. The reason for a higher average space group coverage rate for the latter is that the number of available samples are also lower than the former. However, the average StructureMatcher coverage rates remained higher than or at a similar level at the average space group coverages for these cases (52.38\% and 43.75\%, for formulas with seven and eight polymorphs, respectively). However, it is important to remember that the number of samples for those levels of polymorphism is also limited. For the only unit cell formula with 10 polymorphs, ParetoCSP2 achieved space group match with three out of 10 polymorphs. On the other hand, it achieved StructureMatcher matches with six out of 10 polymorphs. Although the StructureMatcher coverage rate is higher than those with six and eight polymorphs, it is important to note that this result is based on a single sample. Furthermore, the average results for formulas with seven and eight polymorphs are also based on only two available samples. So, they do not represent a reliable estimate of the average coverage rate for compositions with this level of polymorphism.

\FloatBarrier

One of the main challenges in predicting polymorphism lies in effectively sampling and distinguishing closely related structures, particularly for low-symmetry space groups where the configurational space to consider is significantly larger with a greater number of general Wyckoff positions meaning higher degrees of freedom. This can also be noted on the benchmark test set predictions performance section where ParetoCSP2 underperformed for lower-symmetry space group structures. While ParetoCSP2 incorporates a space group-specific optimization criterion, further improvements, such as fine-tuned relaxation strategies or adaptive polymorph-specific constraints, may enhance its ability to recover additional polymorphs. Overall, our results indicate that while ParetoCSP2 demonstrates strong performance in predicting polymorphic structures, while further refinement is necessary to improve its ability to capture highly complex polymorphic systems.

We then extended the application of ParetoCSP2 to generalized polymorphism prediction tasks. For a given chemical formula $(X)$, one can run ParetoCSP2 for $(X)_1$, $(X)_2 \ldots$, $(X)_n$ to cover polymorphism cases with a variable number of atoms in the unit cell with a maximum of $n$ primitive cells. For this, we considered polymorphs of two compositions: Silica (SiO$_2$) and Zinc sulfide (ZnS), which are highly used in industrial and commercial sectors. For Silica, we considered its two highly used polymorphs in industry: $\alpha$-Quartz (unit cell formula: Si$_3$O$_6$, crystal system: trigonal) and Cristobalite (unit cell formula: Si$_4$O$_{8}$, crystal system: cubic). 
Similarly for ZnS, we considered Zincblende (ZnS) (a.k.a. Sphalerite) (unit cell formula: Zn$_4$S$_4$, crystal system: cubic) and Wurtzite (unit cell formula: Zn$_2$S$_2$, crystal system: hexagonal) which are also highly used in various industries. ParetoCSP2 was run with $n = 3$ and $4$ for SiO$_2$ and $n = 2$ and $4$ for ZnS. The results are presented in Fig.~\ref{fig:general_poly}. 

ParetoCSP2 successfully identified both polymorphs of ZnS. Although having more number atoms in the unit cell, ParetoCSP2 successfully predicted the Cristobalite (cubic), but failed to predict the structure of $\alpha$-Quartz (trigonal) for the Silica polymorph cases. This is evident from the results of the benchmark crystal prediction results that ParetoCSP2 has a higher accuracy in predicting structures with higher symmetry crystal systems. Overall, we demonstrated that ParetoCSP2 can be used to cover a wide range of polymorph prediction tasks. It not only identified known polymorphs but also has the potential to predict previously undiscovered crystal structures. These predicted polymorphs, once verified through \textit{ab initio} energy calculations, could be valuable for various industrial applications.

\subsection{Performance evaluation for regular CSP }
Here we compared the ParetoCSP2 algorithm with ParetoCSP~\cite{paretocsp} and GN-OA~\cite{gnoa} to find the most stable structure given a formula. These algorithms are chosen because they are closely related in their approach to leveraging advanced optimization and structure search techniques. We excluded well-established methods such as USPEX~\cite{uspex} and CALYPSO~\cite{calypso} from our comparisons because of their dependence on DFT calculations. Although DFT remains the most accurate approach for computing the energies of crystal structures, its high computational cost significantly limits the efficiency and speed of exploring the crystal configuration space. Furthermore, recent studies have demonstrated that ML IAPs can predict crystal energies several orders of magnitude faster than DFT while maintaining comparable accuracy, making them a viable and efficient alternative for structure prediction. M3GNet~\cite{m3gnet}, an ML IAP covering $89$ elements of the periodic table, and CHGNet~\cite{chgnet}, pre-trained on the Materials Project Trajectory Dataset with $\approx 1.5$ million structures, represent significant advancements in ML-based potentials. These approaches are highly promising for enhancing CSP performance by efficiently predicting formation energies and structural properties, replacing the need for expensive DFT calculations with a small time-accuracy trade-off. A brief summary of each algorithm is presented in Table~\ref{tab:algorithms}.

\begin{table}[!htb]
\small
\centering
\caption{\textbf{A brief summary of compared algorithms used in this work.} Several key differences of ParetoCSP2 with ParetoCSP and GN-OA, especially in terms of space group diversity control, and relaxation frequency show how it improved from its previous similar algorithms.}
\label{tab:algorithms}
\begin{tabular}{|c|c|c|c|}
\hline
\multirow{2}{*}{\textbf{Key factors}} & \multicolumn{3}{c|}{\textbf{Algorithms}} \\ \cline{2-4} 
 & \textbf{GN-OA} & \textbf{ParetoCSP} & \textbf{ParetoCSP2} \\ \hline
\textbf{Search strategy} & BO/PSO & GA + AFPO & GA + AFPO \\ \hline
\textbf{Energy evaluation} & GNN & ML IAP & ML IAP \\ \hline
\textbf{Number of optimization criterion} & $1$ & $2$ & $3$ \\ \hline
\textbf{Population initialization} & Random & Random & PyXtal \\ \hline
\textbf{Space group diversity control} & No & No & Yes \\ \hline
\textbf{Relaxation frequency} & No & Final step only & After every generation \\ \hline
\end{tabular}
\end{table}

In our evaluation, ParetoCSP is used as a baseline for our algorithm improvements, while GN-OA represents a state-of-the-art algorithm for generic crystal structure prediction, making them the most relevant baselines for evaluating the performance enhancements introduced in ParetoCSP2. More specifically, we chose the particle swarm optimization (PSO) version of GN-OA along with the M3GNet IAP due to its higher prediction accuracy in a recent benchmark~\cite{wei2024cspbench} compared to its other two versions. The experimental setups for the experiments are detailed in Supplementary Note S1.3. To ensure a fair comparison and accurately assess the impact of the modifications, we maintained identical experimental settings for ParetoCSP and ParetoCSP2. Furthermore, external hyperparameter optimization can be performed to further enhance the performance of ParetoCSP2.

The average metric values for the chosen 120 crystals across the three evaluated algorithms are presented in Fig.~\ref{fig:metrics_all}, metric values for binary, ternary, and quarternary crystals separately are shown in Supplementary Fig. S4 for better comparison. ParetoCSP2 showed superior performance, outperforming ParetoCSP and GN-OA by a large margin of 53.33\% and 87.04\%, respectively, in terms of the Energy Distance (ED) (binary: 50\% and 70\%, ternary: 70\% and 83.33\%, and quarternary: 29.17\% and 88.82\%, for ParetoCSP and GN-OA respectively). For the Sinkhorn distance (SD), ParetoCSP2 outperformed ParetoCSP and GN-OA by 48.01\% and 59.08\%, respectively (binary: 41.37\% and 52.27\%, ternary: 57.95\% and 63.04\% and quarternary: 40.45\% and 60.15\%, for ParetoCSP and GN-OA, respectively). Similarly, for the Chamfer distance (CD), ParetoCSP2 achieved improvements of 44.81\% and 66.33\% over ParetoCSP and GN-OA, respectively (binary: 45.77\% and 76.17\%, ternary: 48.20\% and 52.71\% and quarternary: 38.78\% and 60.69\%, for ParetoCSP and GN-OA, respectively). In the case of the Hausdorff distance (HD), ParetoCSP2 outperformed ParetoCSP and GN-OA by 50.74\% and 61.61\%, respectively (binary: 46.10\% and 50.24\%, ternary: 55.15\% and 64.71\% and quarternary: 49.57\% and 66.97\%, for ParetoCSP and GN-OA, respectively). Finally, for the Fingerprint distance (FP), ParetoCSP2 surpassed ParetoCSP and GN-OA by 46\% and 55.98\%, respectively (binary: 45.08\% and 64.92\%, ternary: 58.02\% and 60.23\% and quarternary: 32.41\% and 37\%, for ParetoCSP and GN-OA, respectively). These results clearly highlighted the effectiveness of incorporating the distribution of space groups as an additional optimization criterion alongside final energy and genotypic age.

\begin{figure}[!htb]
    \centering
    \includegraphics[width=1\linewidth]{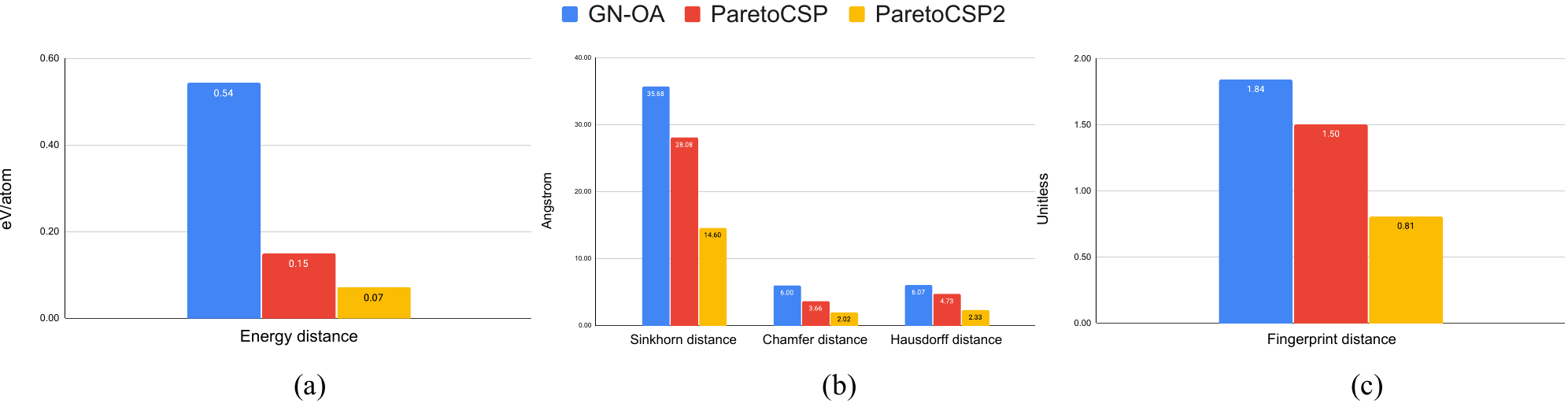}
    \caption{\textbf{Performance metrics' comparison of ParetoCSP2 vs ParetoCSP and GN-OA for all benchmark set crystals.} The plots are divided by type of unit for better comparison: (a) ED, (b) SD, CD, and HD, and (c) FP. ParetoCSP2 achieved performance improvements ranging from 44.8\% to 87.04\% over the other two algorithms, demonstrating its effectiveness in general CSP tasks.}
    \label{fig:metrics_all}
\end{figure}

\begin{figure}[!htb]
    \centering
    \includegraphics[width=1\linewidth]{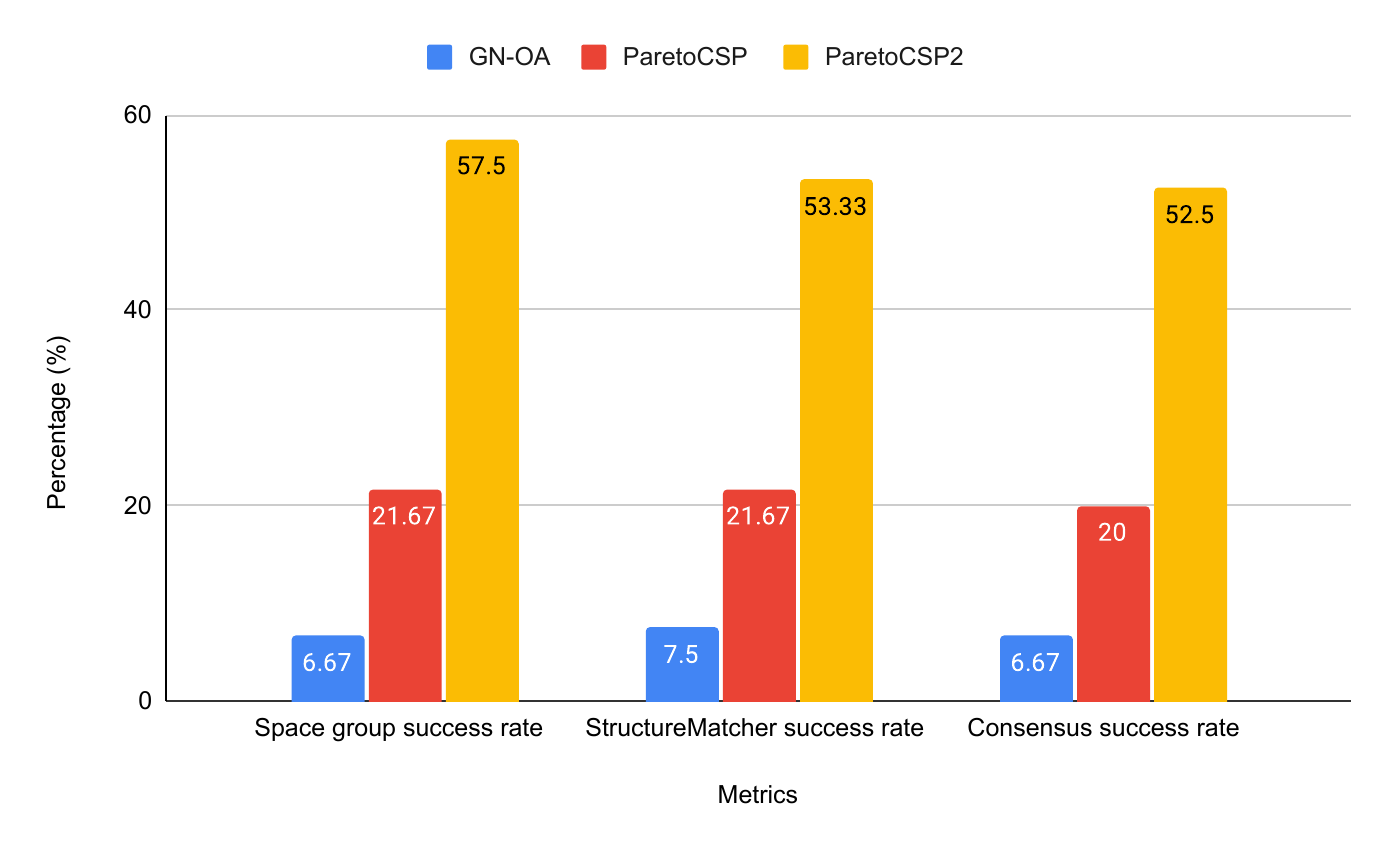}
    \caption{\textbf{Space group and StructureMatcher success rate of ParetoCSP2 vs ParetoCSP and GN-OA.} ParetoCSP2 outperformed ParetoCSP (GN-OA) by a factor of 2.65 (8.62), 2.46 (7.11), and 2.63 (7.87) for space group success rate, StructureMatcher success rate, and consensus success rate, respectively. This difference in performance demonstrates the effectiveness of ParetoCSP2's new optimization criterion for adaptive space group control and highlights the importance of incorporating structural relaxation after each generation.}
    \label{fig:success_rate}
\end{figure}

Fig.~\ref{fig:success_rate} shows the comparison of the space group match rate and the Pymatgen StructureMatcher match rate of ParetoCSP2 with ParetoCSP and GN-OA. Notably, unlike prior research, we observed that the StructureMatcher can incorrectly identify two similar structures as identical despite differences in their space groups. Consequently, we used success rates in conjunction with the space group match rate to ensure a more comprehensive evaluation of the predicted structures' performance. The space group diversity mechanism in ParetoCSP2 enabled a considerable improvement of 165.34\% in the space group match rate over that of ParetoCSP, while GN-OA underperformed it by a staggering 762.07\%. For the Pymatgen StructureMatcher rate, ParetoCSP2 again outperformed ParetoCSP and GN-OA by 146.10\% and 611.07\%, respectively. Relaxation of all structures after each generation of ParetoCSP2 allows it to work with more refined structures each time, while a better initialization technique allows it to work with more valid structures from the beginning, leading it to predict considerable quality structures. Evidently, ParetoCSP outperformed ParetoCSP and GN-OA by 162.50\% and 687.11\%, respectively, considering both the space group and the StructureMatcher match rate. We presented the three algorithms' success for each crystal in terms of space group, StructureMatcher, and consensus match in Supplementary Tables S3, S4, and S5, respectively.

Supplementary Fig. S5 presents some cases using the VESTA tool~\cite{vesta}, where ParetoCSP2 is shown to be able to successfully identify the optimal crystal, but the other two algorithms failed. Due to the absence of an efficient space group exploration strategy, ParetoCSP and GN-OA were unable to predict even high-symmetry structures, such as Nb$_3$Si (cubic), which has fewer configurational degrees of freedom. This challenge was compounded even more for lower-symmetry structures, such as Ca(CdP)$_2$ (trigonal) and CeCr$_2$Si$_2$C (tetragonal).

\FloatBarrier

\FloatBarrier

\subsection{Comparative analysis of space group diversity}
\begin{figure}[!htb] 
    \centering
    \begin{minipage}[c]{0.495\textwidth}
        \centering
        \includegraphics[width=\textwidth]{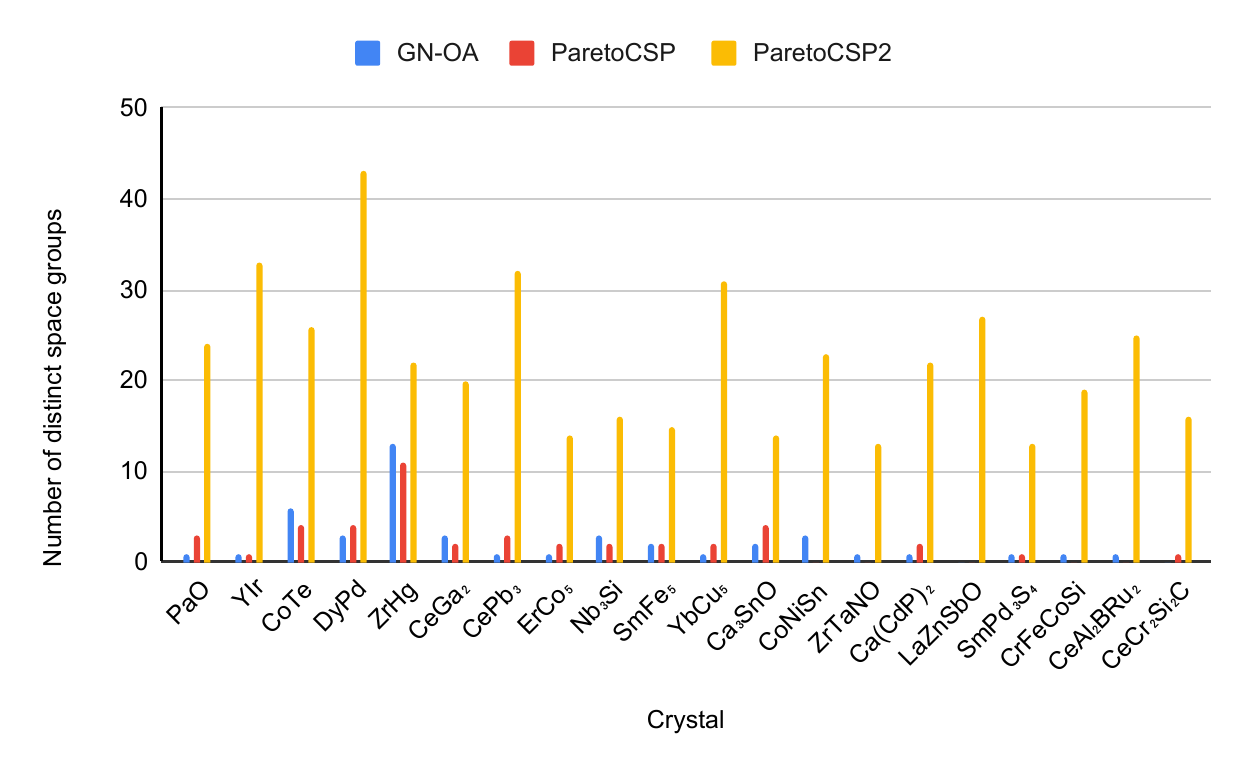}
        \subcaption{}
        \label{fig:sg_50}
    \end{minipage}
    \begin{minipage}[c]{0.495\textwidth}
        \centering
        \includegraphics[width=\textwidth]{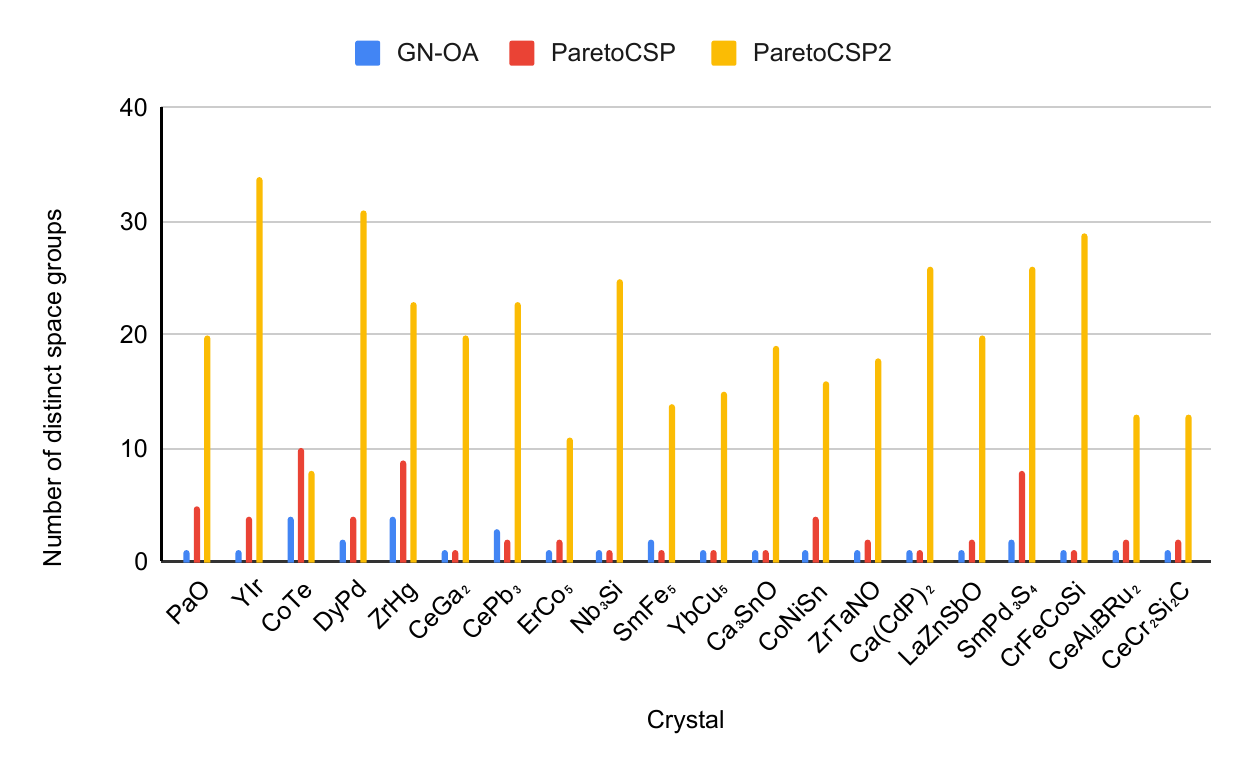}
        \subcaption{}
        \label{fig:sg_500}
    \end{minipage}
     \caption{\textbf{Space group diversity ($\uparrow$) of ParetoCSP2 vs ParetoCSP and GN-OA for a population size of 100 for a set of sample crystal compositions.} Space group diversity is measured by the number of distinct space group individuals (structures) existing in the population. The space group diversity is shown at two stages: (a) an early stage of the algorithm (50 generations here) and (b) after termination (500 generations here).  We observed that ParetoCSP2 constantly maintained a much larger space group diversity compared to the other two algorithms at both stages, highlighting the effectiveness of its new optimization criterion.}
     \label{fig:sg_diversity}
\end{figure}

To understand how the space group exploration strategy contributes to the effectiveness of ParetoCSP2, we check the space group diversity of the population during its search process. Fig.~\ref{fig:sg_diversity} presents the number of distinct space groups in both the early stage (after 50 generations) and at termination ($\mathcal{G}$ generations, where $\mathcal{G}$ is typically set to 250 to 1000 here) for a set of sample crystal compositions, using a population size of 100. The box plot for this set of crystals is shown in Supplementary Fig. S6 to visually represent the distribution for a better comparison. The results indicate that ParetoCSP2 consistently achieved significantly higher space group diversity compared to ParetoCSP and GN-OA at both stages. Specifically, ParetoCSP2 frequently generated structures with more than 25-30 distinct space groups, whereas ParetoCSP and GN-OA typically produced structures that span fewer than five distinct space groups due to their fast structural convergence. This demonstrates the effectiveness of ParetoCSP2's new space group-specific independent optimization criterion in its multi-objective genetic algorithm. Moreover, it was found that candidate structures with different space groups generated by ParetoCSP and GN-OA often converge to the same structure after relaxation, further reducing the number of distinct space groups in the population, complicating polymorph prediction (see Subsection ``Performance evaluation for polymorphism prediction'' for details). However, the relaxation of all structures after each generation in ParetoCSP2 eliminates this problem, preserving a truly diverse population set after each generation.

\FloatBarrier

We took some case studies of sample non-polymorphic compositions (e.g., Ca(CdP)$_2$) and polymorphic compositions (e.g., Nb$_3$Si) and plotted the progression of space group diversity for the three algorithms. Two of them are shown in Fig.~\ref{fig:sg_diversity_progression} and the rest are shown in Supplementary Fig. S7. We noticed that ParetoCSP2 consistently maintained significantly higher space group diversity throughout the optimization process, whereas ParetoCSP and GN-OA remained restricted to a much smaller number of distinct space groups. In all cases, ParetoCSP2 rapidly increased the space group diversity within the first 100 to 150 generations before stabilizing at a significantly higher level than the other two algorithms in the majority cases. However, there are a few cases such as GaP and CeAl$_2$BRu$_2$ where the stability point is closer to that of ParetoCSP and GN-OA, but still maintained a higher overall space group diversity than them. In contrast, ParetoCSP and GN-OA maintain consistently low space group diversity throughout the evolutionary process, suggesting their inadequate techniques for diverse space group exploration.

The quadratic fitted trend lines further emphasize the effectiveness of ParetoCSP2’s adaptive space group control mechanism, which allows it to efficiently explore a much broader range of space groups, even for complex polymorphic structures, such as of Li$_2$NiO$_2$. Although we observed that the space group diversity of ParetoCSP2 slightly decreased after around 200 generations in most cases, it remained significantly higher than that of ParetoCSP and GN-OA. This indicates that ParetoCSP2 effectively preserves structural diversity despite the inherent tendency of genetic algorithms to converge toward lower-energy solutions. These findings highlight the advantages of ParetoCSP2’s space group-specific optimization criterion and iterative relaxation strategy to maintain diversity, allowing for a more thorough exploration of the crystal structure space.
Overall, we found that by minimizing the maximum number of structures for each space group, ParetoCSP2 achieved a diverse representation of space group structures within its population, leading to its superior performance for polymorphism CSP and regular CSP.

\begin{figure}[!htb] 
    \centering
    \begin{minipage}[c]{0.495\textwidth}
        \centering
        \includegraphics[width=\textwidth]{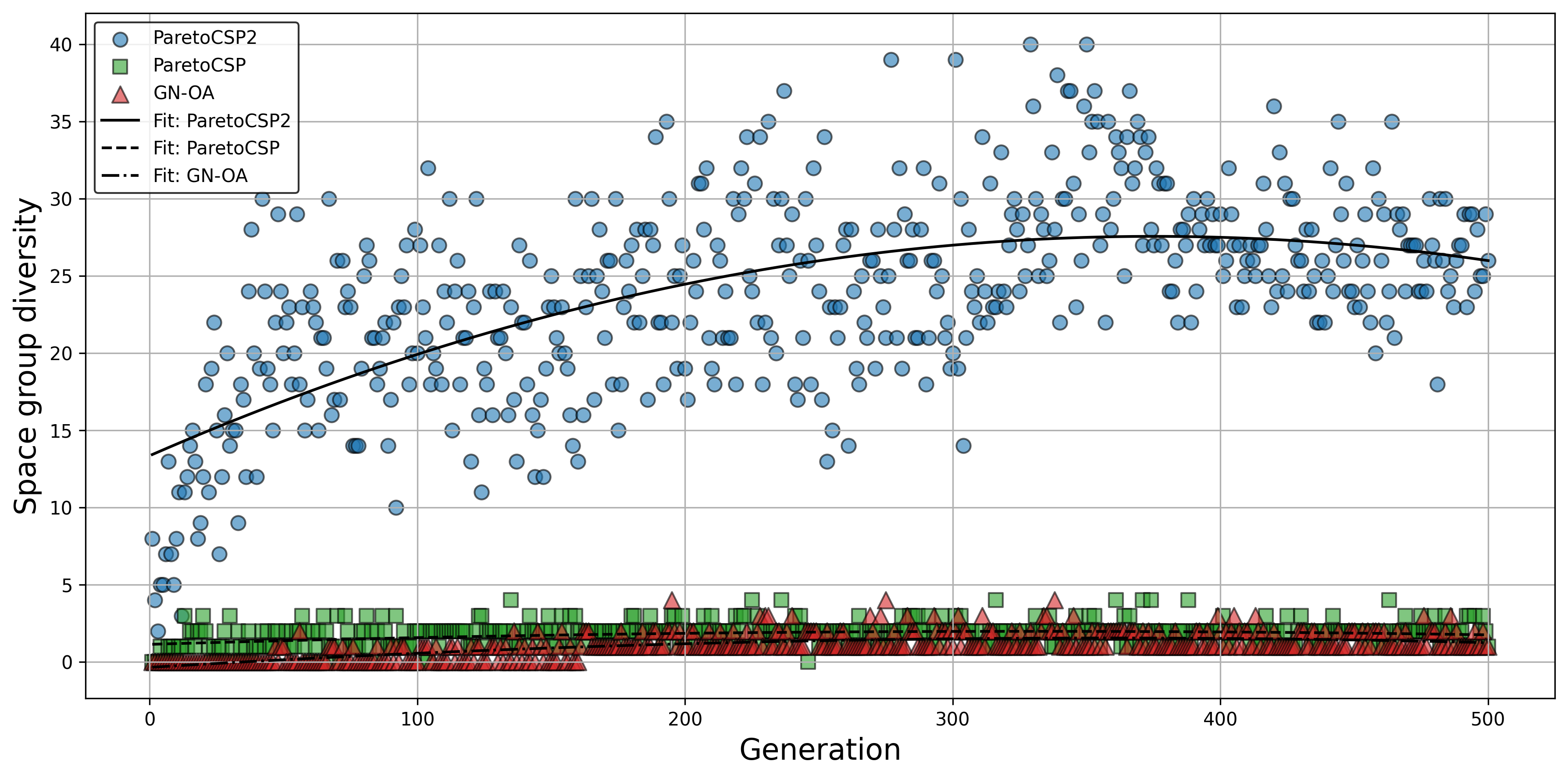}
        \subcaption{}
        \label{fig:sg_1}
    \end{minipage}
    \begin{minipage}[c]{0.495\textwidth}
        \centering
        \includegraphics[width=\textwidth]{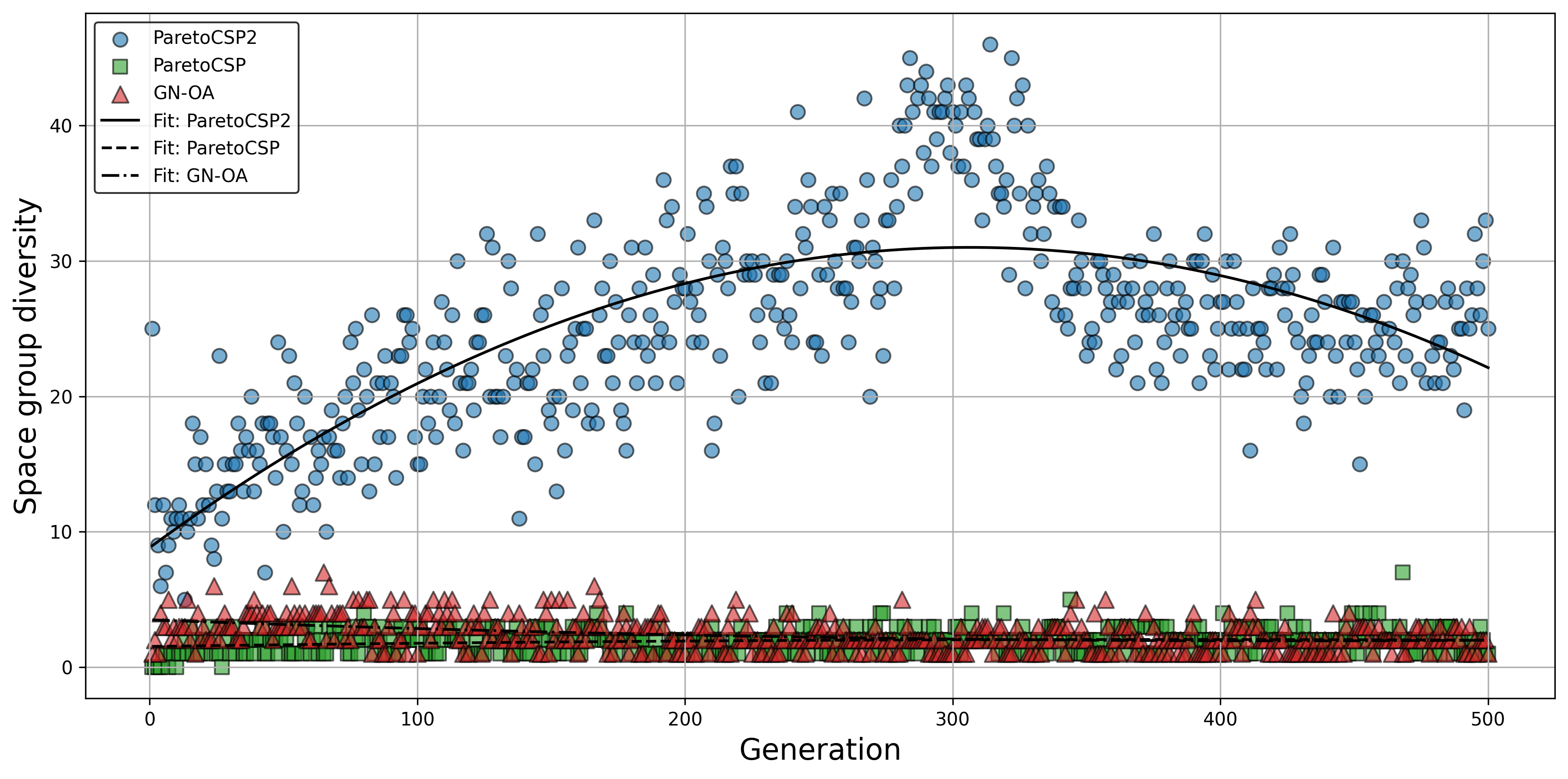}
        \subcaption{}
        \label{fig:sg_2}
    \end{minipage}
     \caption{\textbf{Space group diversity ($\uparrow$) progression of ParetoCSP2 vs ParetoCSP and GN-OA for two case studies for a population size of 100.} (a) Ca(CdP)$_2$(non-polymorphic case) and (b) Nb$_3$Si (polymorphic case) show the evolution of space group diversity over 500 generations. ParetoCSP2 consistently achieves significantly higher space group diversity throughout the optimization process, while ParetoCSP and GN-OA remain limited to fewer distinct space groups. The trend lines indicate the quadratic fitted progression of space group diversity for each algorithm, highlighting ParetoCSP2's superior exploration capability and adaptability in both non-polymorphic and polymorphic cases.}
     \label{fig:sg_diversity_progression}
\end{figure}

\FloatBarrier

\subsection{Comparative analysis of valid structure generation}
An invalid crystal is a generated structure that does not follow basic physical or chemical rules, such as atoms being unrealistically close or too far apart, making them unstable. Generating a high number of valid structures is a key objective for any CSP algorithm. However, prior methods such as GN-OA and ParetoCSP often produce very few valid candidates, significantly reducing their ability to effectively explore the search space and identify global minima.

\begin{figure}[!htb] 
    \centering
    \begin{minipage}[c]{0.495\textwidth}
        \centering
        \includegraphics[width=\textwidth]{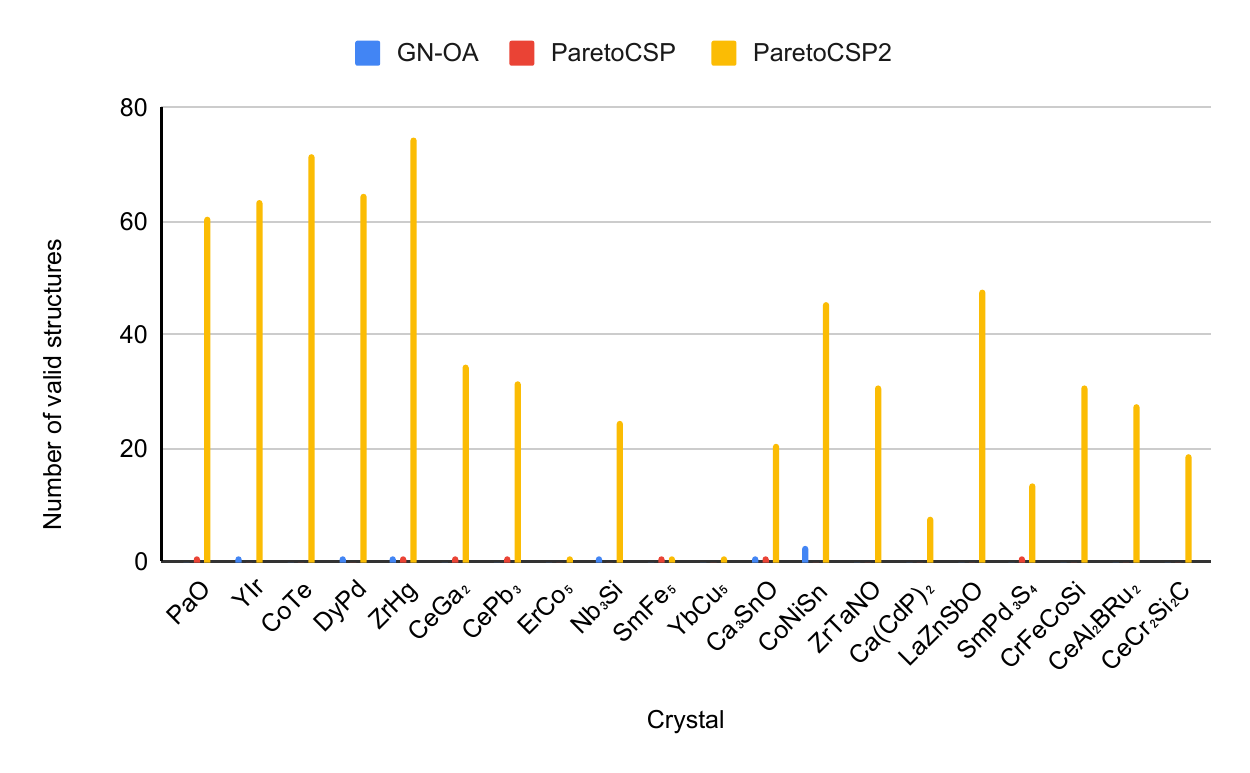}
        \subcaption{}
        \label{fig:valid_init}
    \end{minipage}
    \begin{minipage}[c]{0.495\textwidth}
        \centering
        \includegraphics[width=\textwidth]{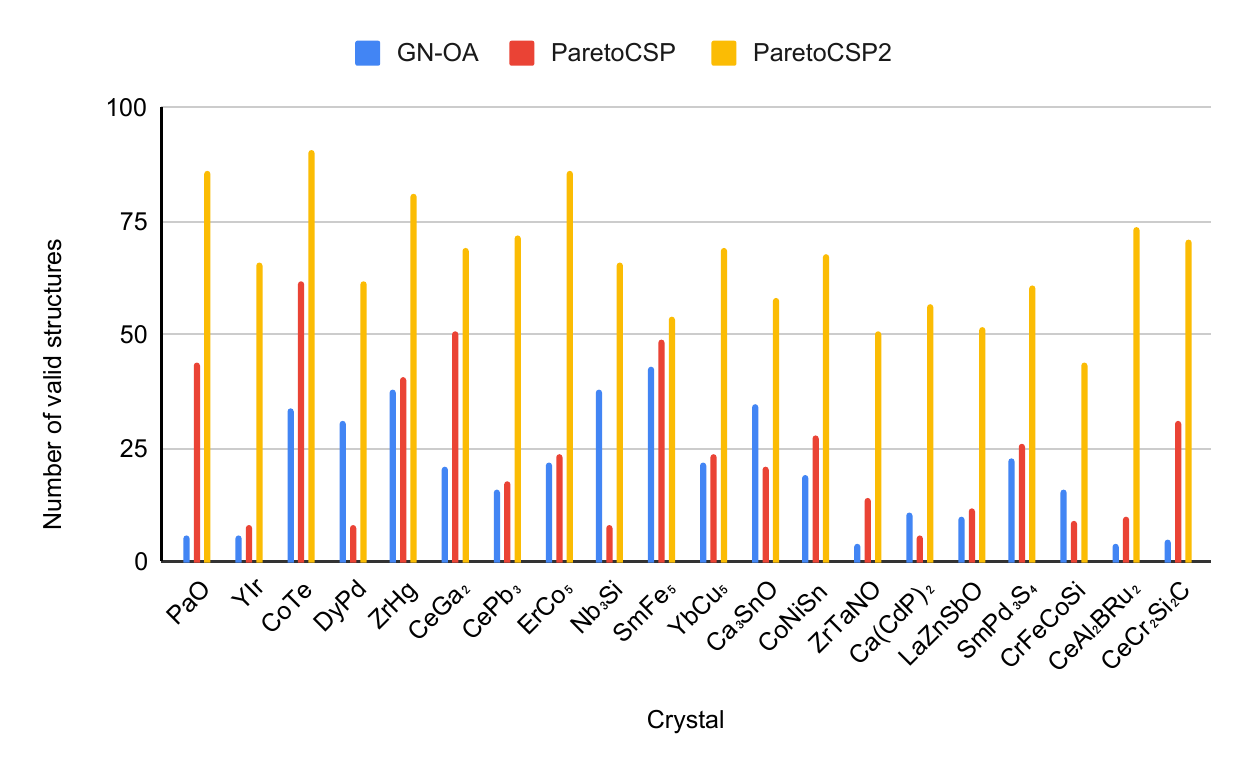}
        \subcaption{}
        \label{fig:valid_term}
    \end{minipage}
     \caption{\textbf{Comparison of the number of valid structures generated ($\uparrow$) by ParetoCSP2 vs ParetoCSP and GN-OA for a population size of 100 for a set of sample crystal compositions.} The number of valid structures in the population is shown at two stages: (a) after the first generation, where ParetoCSP2 utilizes PyXtal-based initialization, while ParetoCSP and GN-OA rely on random initialization, resulting in a significantly higher count of valid structures in ParetoCSP2, and (b) after the termination of the algorithms. Although the number of valid structures increased on average for all three algorithms, we observed that ParetoCSP2 maintained a larger valid structure count than the other two. The combination of PyXtal-based initialization and the new space group-specific optimization criterion in ParetoCSP2 facilitates better space group exploration, leading to the discovery of a more diverse and valid set of structures throughout the evolutionary process.}
     \label{fig:valid}
\end{figure}

Replacing the random population initialization method with an efficient method from the \texttt{PyXtal}~\cite{pyxtal} library, ParetoCSP generated more valid structures from the beginning. This improvement allowed subsequent generations of the algorithm to operate with a larger pool of valid solutions, ultimately maintaining a consistently higher count of valid structures. Furthermore, PyXtal facilitated the generation of structures from diverse space groups during the initialization phase, allowing ParetoCSP2 to maintain a diverse set of space group structures from the outset. The results are reflected in Fig.~\ref{fig:valid_init}, where the valid structure count for the three algorithms after the first generation for a population size of 100 is shown. We noticed a significant difference in the number of valid structures of ParetoCSP2 compared to ParetoCSP and GN-OA. We observed that a randomly initialized population often times does not even have a single valid crystal to begin with, whereas PyXtal generated populations frequently have more than 50 valid crystals, which helps ParetoCSP2's genetic operators to work with a better pool of population from the beginning. This demonstrates the major limitation of using a randomly generated initial population.

Combined with the separate space group-wise count optimization criterion, this approach served as a foundation for ParetoCSP2 to effectively explore the diverse space groups for each composition and maintain greater population diversity compared to the other two algorithms throughout all generations. The number of valid crystals in the population after the termination of the algorithms is shown in Fig.~\ref{fig:valid_term}. The box plot for this set of crystals is shown in Supplementary Fig. S8 to visually represent the distribution for a better comparison of valid structure counts in both an early stage and termination of the algorithm. Although we observed an increase in total valid structures for all three algorithms since the beginning, ParetoCSP2 has a much larger number of valid structures compared to the other two algorithms. This shows the importance of effective space group exploration in generating more valid solutions, thereby guiding the algorithm toward the global minimum. Moreover, most of the valid structures generated by ParetoCSP and GN-OA basically converge to a very few (often the same) space groups after a final relaxation step. This highlights the importance of ParetoCSP2's relaxation after each generation, which helps prevent this situation.

We plotted the progression of valid structure count in a population set by taking some case studies of non-polymorphic compositions (e.g., CaCd$_2$P$_2$) and polymorphic compositions (e.g., Nb$_3$Si). Two of them are shown in Fig.~\ref{fig:valid_structures_progression} and the rest are shown in Supplementary Fig. S9. ParetoCSP2 consistently maintained a significantly higher count of valid structures than the other two algorithms from the very beginning. The increase in total valid structures in the population followed a similar trend in both the polymorphic and non-polymorphic case for ParetoCSP2 - in both cases the number of valid structures gradually increased from generation 1 to around generation 100 and then either became stabilized or a rather faster increase after 300 generations, often achieving 70-80 valid individuals among the 100-sized population. We found that ParetoCSP2's stabilization points are considerably higher than the other two algorithms. However, we did not find any specific trend among GN-OA and ParetoCSP regarding which algorithm performed comparatively better for valid structure generation.

\begin{figure}[!htb] 
    \centering
    \begin{minipage}[c]{0.495\textwidth}
        \centering
        \includegraphics[width=\textwidth]{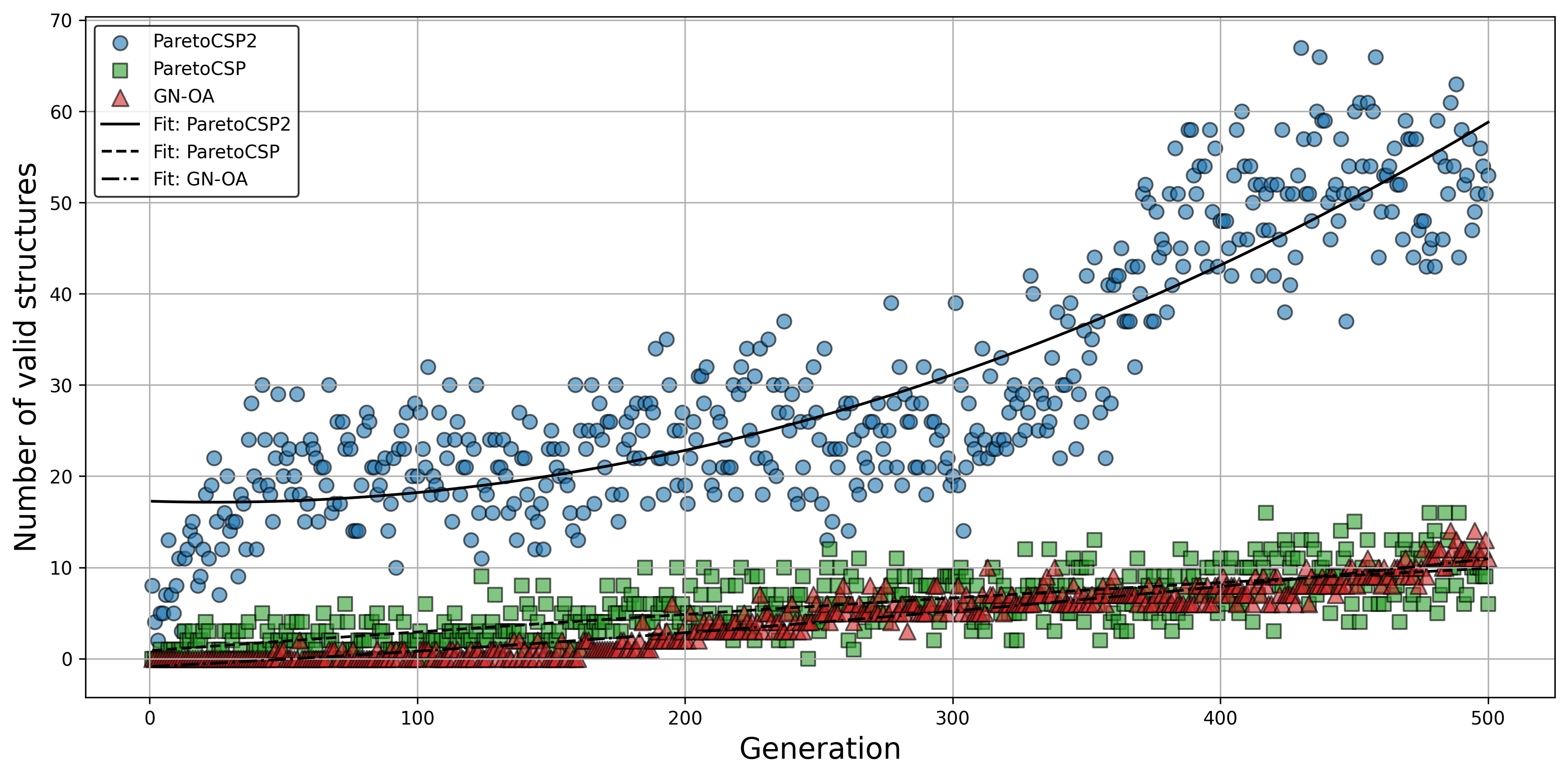}
        \subcaption{}
        \label{fig:valid_1}
    \end{minipage}
    \begin{minipage}[c]{0.495\textwidth}
        \centering
        \includegraphics[width=\textwidth]{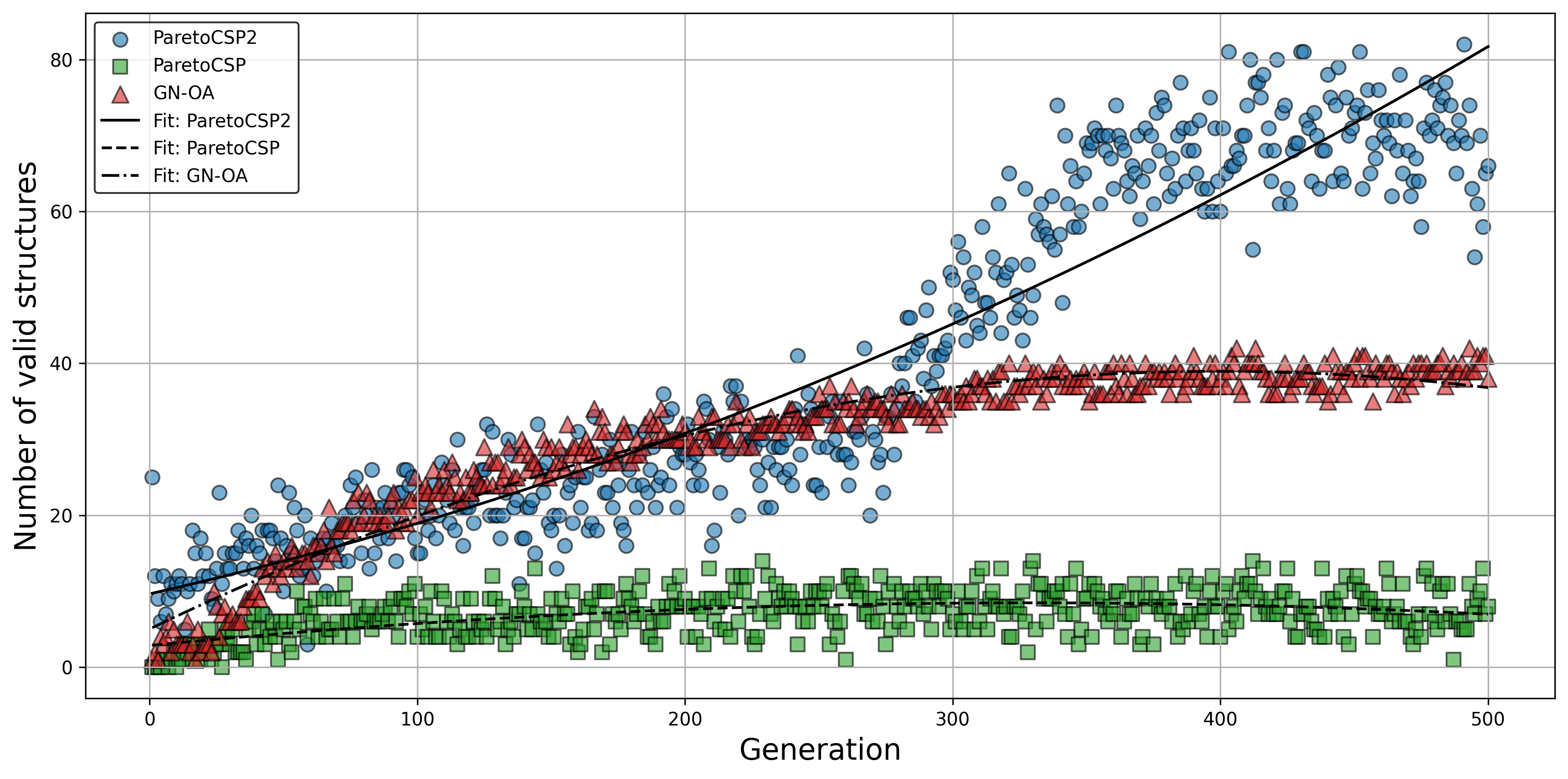}
        \subcaption{}
        \label{fig:valid_2}
    \end{minipage}
     \caption{\textbf{Valid structure count ($\uparrow$) progression of ParetoCSP2 vs ParetoCSP and GN-OA for two case studies for a population size of 100.} (a) Ca(CdP)$_2$(non-polymorphic case) and (b) Nb$_3$Si (polymorphic case) show the evolution of valid structure count over 500 generations. In both cases, all three algorithms achieved increasing number of valid structures, but the increase in ParetoCSP2 is significantly higher than that of ParetoCSP and GN-OA. The trend lines represent the quadratic fitted progression of valid structure counts, demonstrating the effectiveness of ParetoCSP2's PyXtal-based initialization and space group-specific optimization criterion in facilitating better exploration and preservation of valid structures over generations.}
     \label{fig:valid_structures_progression}
\end{figure}

\subsection{Comparative analysis of convergence speed}
Convergence speed is important in CSP because the search space is vast and computational resources are limited in most cases. Faster convergence allows the algorithm to identify low-energy structures efficiently, reducing the time and cost required for large-scale or high-throughput predictions.

ParetoCSP2's advanced approach to explore diverse space groups, combined with the relaxation of all structures after each generation, enables it to identify the optimal structure significantly faster, often within 1–10 generations. In contrast, similar algorithms such as ParetoCSP and GN-OA typically achieve the energetically optimal structure only after relaxing the lowest-energy structure at the termination of the algorithm. By performing relaxation at every generation, ParetoCSP2 operates with already refined structures, allowing it to produce solutions that are not only more accurate but also faster compared to algorithms that delay relaxation until the end. This iterative refinement improves the quality of the structures at each step of the genetic operations and accelerates convergence to the optimal solution.

\begin{figure}[!htb]
    \centering
    \includegraphics[width=1\linewidth]{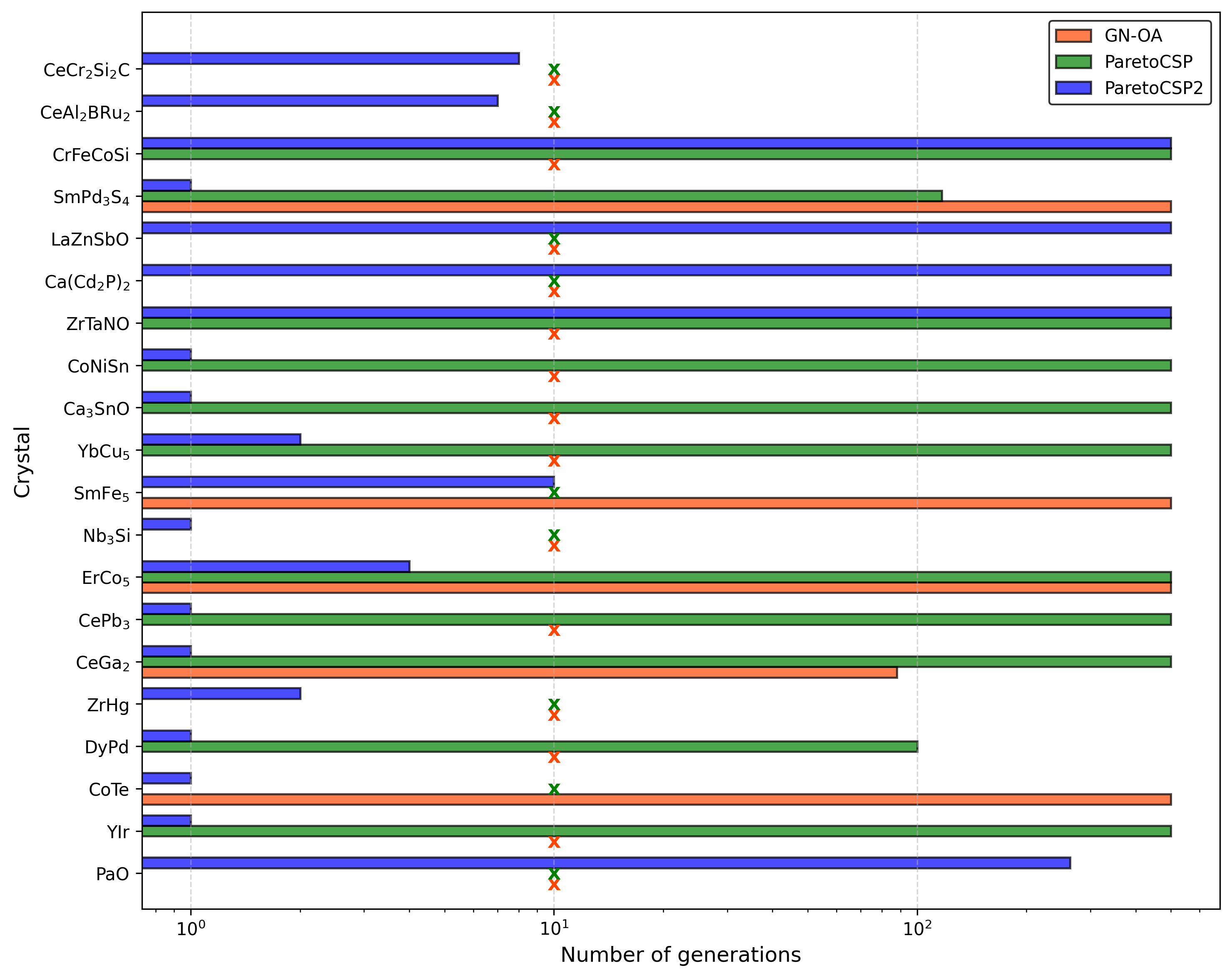}
    \caption{\textbf{Convergence speed ($\downarrow$) of ParetoCSP2 vs ParetoCSP and GN-OA for a population size of 100 for a set of sample crystal compositions}. The $x$-axis represents the number of generations required for each algorithm to achieve the optimal prediction, with a ``\xmark'' indicating cases where an algorithm did not converge to the optimal solution. ParetoCSP2 achieved most optimal prediction within 1-10 generations, which is significantly faster than the other two algorithms, demonstrating the need for iterative relaxation for faster convergence to achieve the refined structure in an earlier stage of the algorithm.}
    \label{fig:convergence}
\end{figure}

Fig.~\ref{fig:convergence} shows the convergence speed for the three algorithms for a set of sample crystal compositions for a population size of 100. A cross mark (\xmark) is used if an algorithm failed to provide the optimal solution. We chose the number of generations taken to converge to the optimal solution as the speed metric. We found that ParetoCSP2 successfully identified the correct structures in significantly fewer generations than ParetoCSP and GN-OA. We also chose the log scale in Fig.~\ref{fig:convergence} to highlight the huge differences of ParetoCSP2 compared to ParetoCSP and GN-OA in terms of convergence speed. This suggests the critical importance of incorporating structural relaxation during the early stages of the algorithm, instead of performing it just as a final refinement step, for improved performance and faster convergence in similar CSP algorithms.

\subsection{Failure case study}
Despite its improved performance, ParetoCSP2 still failed to predict certain crystal structures. Here, we aim to explore the underlying factors contributing to these failures and its remaining limitations, especially in space group coverage, to better understand its prediction boundaries.

\begin{figure}[!htb]
    \centering
    \includegraphics[width=1\linewidth]{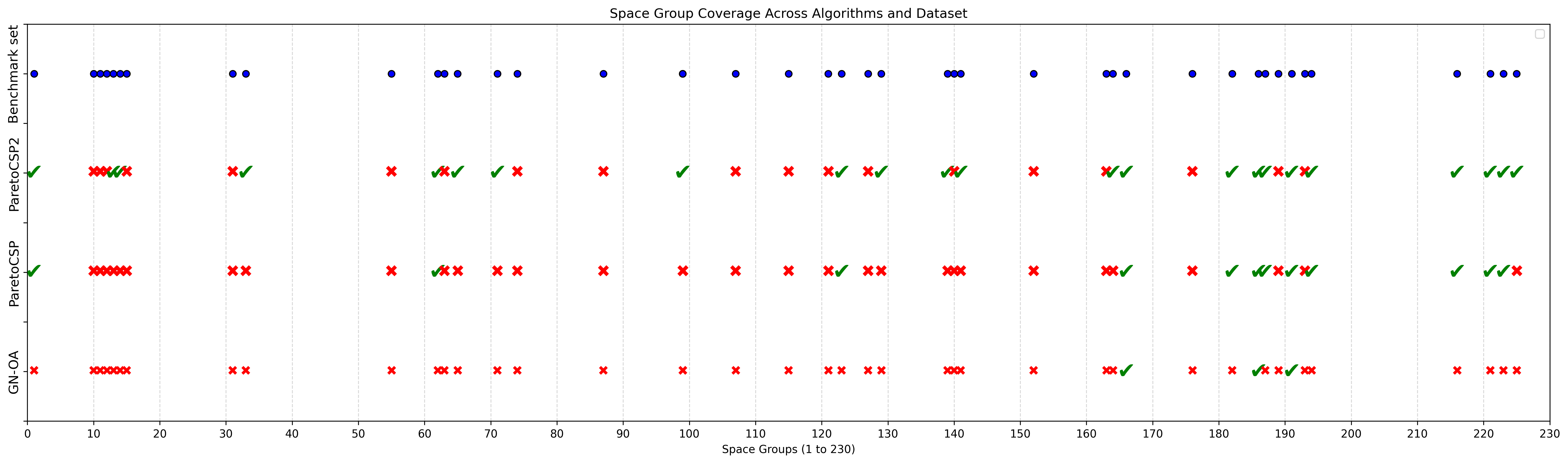}
    \caption{\textbf{Comparison of the space group coverage achieved by GN-OA, ParetoCSP, and ParetoCSP2 vs the full set of space groups present in the benchmark set.} Each row corresponds to an algorithm (or the benchmark set), and each column represents a distinct space group (from 1 to 230). For each space group, a blue dot ($\cdot$) indicates the presence of crystals with that space group in the benchmark set, a green checkmark ($\cmark$) indicates that an algorithm successfully predicted a crystal from the benchmark set with that space group, and a red cross (\xmark) indicates that an algorithm failed to predict any crystal with that space groups. ParetoCSP2 demonstrated significantly broader coverage, successfully recovering many more space groups than both GN-OA and ParetoCSP, highlighting its effectiveness in generating structurally diverse predictions. However, its prediction performance still lacked for certain space group structures, especially those with lower symmetry structures, suggesting room for further improvement in exploration strategies.}
    \label{fig:failure}
\end{figure}

Fig.~\ref{fig:failure} presents the coverage of the space groups for the structures each algorithm successfully predicted or failed among the benchmark set. The test set contains structures spanning a wide range of space groups, indicated by blue dots ($\cdot$). The green checkmarks ($\cmark$) denote space groups for which an algorithm successfully predicted at least one structure of that space group, while the red crosses (\xmark) represent the opposite. The figure shows that ParetoCSP2 demonstrates superior space group coverage compared to ParetoCSP and GN-OA, successfully predicting structures spanning a broader range of space groups. In particular, ParetoCSP2 performs particularly well for high-symmetry space groups (lower configurational complexity and fewer structural degrees of freedom) where other algorithms underperform. This success can be attributed to ParetoCSP2's iterative relaxation mechanism and space group-specific optimization, which help maintain structural diversity throughout the evolutionary process and prevent premature convergence to common low-symmetry space groups.

Despite its advantages, ParetoCSP2 still exhibits failure cases, particularly in low-symmetry space groups (e.g., monoclinic and orthorhombic structures with higher space group numbers), where its predictions remain incomplete or inaccurate. Some failure cases of ParetoCSP2 are shown in Supplementary Fig. S10. Low-symmetry space groups have fewer symmetry operations, which results in a greater number of general Wyckoff positions. These general positions have more free parameters, meaning each atom can vary more independently in the 3D space. Consequently, the number of possible atomic arrangements increases drastically, expanding the search space and making structure prediction significantly more difficult. Moreover, the lack of symmetry in low-symmetry structures limits the ability of algorithms to leverage symmetry constraints for pruning or guiding the search, further increasing the computational burden and the likelihood of generating invalid or energetically infeasible structures. This makes it challenging for genetic algorithms to efficiently converge toward the correct solution, as the search space is not only larger, but also lacks the guidance provided by symmetry constraints. Additionally, low-symmetry structures tend to have multiple metastable configurations that are close in energy, making it difficult to identify the true ground-state structure without extensive relaxation and validation steps.

This difficulty is further increased by the limitations of ML IAP models such as M3GNet or CHGNet, which may not effectively capture fine energy differences among structurally similar low-symmetry configurations. Although these models are trained on a vast amount of materials (both stable and unstable), they are not DFT-level accurate. For this reason, during the search process, it often assigns a generated crystal a wrong energy value, or even sometimes an energy value lower than it assigns to the ground truth structure, which severely misguides the algorithm for the exact structure prediction. Supplementary Fig. S11 shows a case of ZrGa, where it assigned an intermediate structure an energy value greater than that of its most stable form, and in the process misguided the algorithm to predict the exact structure. This shows the need to develop more accurate ML IAPs to implement better CSP algorithms.

\section{Discussions}
We presented ParetoCSP2, an improved CSP algorithm designed for polymorphism prediction given a chemical composition. Improving on our previous ParetoCSP framework (which was not originally developed for polymorphism prediction), ParetoCSP2 incorporates several key improvements that enhance both structural diversity and predictive accuracy, two critical requirements for effective polymorph exploration. Traditional CSP methods often focus on finding a single lowest energy structure, often overlooking the existence of multiple distinct yet energetically comparable structures (i.e., polymorphs). Moreover, a key limitation of traditional CSP algorithms is that structures belonging to different space groups often converge to the same final configuration after a sufficiently large number of structural relaxation steps~\cite{hessmann2025accelerating}, making them unsuitable for polymorphism prediction tasks. In this work, we addressed these limitations by introducing techniques that not only promote the generation of structurally diverse candidate solutions, but also guide the search process to preserve symmetry and distinguish between energetically competitive polymorphs throughout the optimization pipeline.

First, we introduced a key enhancement in ParetoCSP2 by penalizing over-representation of any single space group in the population through a new optimization criterion alongside the energy and the genotypic age. By minimizing the maximum count of individuals belonging to the same space group in the population, ParetoCSP2 achieved significantly higher space group diversity adaptively throughout the search process, which is different from the heuristic approach reported in \cite{spai}. This is crucial for polymorphism CSP, where the ability to explore diverse symmetry configurations often determines the success of identifying alternative stable structures. Our results clearly show that ParetoCSP2 maintained more than six times the space group diversity on average compared to ParetoCSP and GN-OA, even after hundreds of generations.

Second, we noticed that, despite generating a diverse population, most of the crystals generated by the algorithm are invalid (physically impossible to form). We used a PyXtal-based population initialization to significantly remove this issue. Unlike random initialization, which often generates invalid or high-energy structures, PyXtal generates symmetry-consistent structures with realistic atomic positions. This significantly increases the number of valid structures from the very first generation, providing a more suitable population for genetic operations. Our experiments showed that ParetoCSP2 frequently began with more than 50 valid structures in a population of 100, whereas random methods often started with fewer than 10. This early advantage propagates throughout the optimization, enabling faster convergence and better final predictions.

Third, to reliably predict all polymorphic forms of a given chemical formula, it is important to keep track of lower energy structures of multiple space groups. This challenge is compounded by the fact that many structures converge to similar configurations after sufficient relaxation steps, as previously mentioned. We overcame this issue by relaxing all the structures at each generation. This ensures that the tracked structures of various space groups will not converge to the same space group structures as they are already relaxed. In contrast, most algorithms perform relaxation only at the final step due to the computational cost, which limits their prediction performance. We handle this issue by using efficient implementation techniques, such as batching and garbage collection. Additionally, we performed shallow relaxations after each generation and deep relaxation at the end to decrease prediction time. Our results demonstrated that ParetoCSP2 typically found the correct structure within the first 10 generations, whereas other algorithms typically found the correct structure after structure relaxation at the end of the algorithm. This showed the potential of ParetoCSP2 in high-throughput structure prediction for various tasks.

Finally, we should note that polymorphisms are difficult to define for inorganic crystals because of their periodic patterns and flexibility of atomic arrangements in crystalline lattices. Moreover, crystal formations are influenced by other factors such as temperature, pressure, and synthesis conditions. In contrast, organic molecules are generally limited by fixed intramolecular bonding patterns and have limited flexibility in their packing arrangements, resulting in fewer polymorphic forms~\cite{yao2022recent,cruz2020open}. For instance, carbon exists as both diamond, with a three-dimensional tetrahedral lattice, and graphite, with planar hexagonal layers—both composed solely of carbon atoms. Because each polymorph has a unique packing arrangement and symmetry, they often differ in the number of atoms contained in their unit cell. As a CSP algorithm typically requires the exact number of atoms in the unit cell, we consider a special case of polymorphism structure prediction problem in this work: two structures are considered polymorphs if they have the same chemical formula and the same number of atoms in the unit cell. So, for simplicity and as a pioneering work in crystal polymorph prediction, ParetoCSP2 does not directly handle cases, such as SiO$_2$ polymorphs like $\alpha$-Quartz (unit cell formula: Si$_3$O$_6$, crystal system: trigonal) and Cristobalite (unit cell formula: Si$_4$O$_{8}$, crystal system: cubic) simultaneously, but directly handle cases such as $\alpha$-Quartz (unit cell formula: Si$_3$O$_6$, crystal system: trigonal) and $\beta$-Quartz (unit cell formula: Si$_3$O$_6$, crystal system: hexagonal) simultaneously. However, ParetoCSP2 can be adopted to make predictions on multiple units for a chemical formula. So, given a chemical formula $X$, we can make polymorphism predictions for $(X)_1$, $(X)_2$, $\ldots$, $(X)_n$ given a unit value $n$ to cover cases like $\alpha$-Quartz and Cristobalite simultaneously also (see Fig.~\ref{fig:general_poly}).

In general, our algorithm demonstrated strong accuracy in predicting polymorphs for various chemical compositions. It achieved good performance for formulas with 2-3 different polymorphs, with performance gradually declining as the number of possible polymorphs increased. However, it still identified more than half of the polymorphs in most cases, which is also a positive outcome, since only a small subset of polymorphs are typically significant in various applications. For example, among the 321 polymorphs of SiO$_2$ and 145 polymorphs of ZnS reported in the MaterialsProject database~\cite{10.1063/1.4812323}, only 3-10 polymorphs are considered industrially important, and so are highly researched. So, even low coverage rates of polymorphism prediction can still be significant because the predicted polymorphs might be either important in different practical applications or the polymorph might be an exceptionally novel material.

The decrease in accuracy with increasing polymorphs suggests that ParetoCSP2 may benefit from additional refinement in polymorph screening, such as in distinguishing subtle structural variations within the same space group. Moreover, a sequential approach to polymorph prediction can be devised, in which the algorithm focuses on predicting one or a few polymorphs at a time, rather than attempting to identify all possible polymorphic forms simultaneously. This iterative strategy, may help narrow down the search gradually and improve the likelihood of discovering additional polymorphs in subsequent prediction phases. Also, preparing a better test set with more diverse polymorphs from multiple databases where each class of polymorphs has a similar number of samples could provide a more accurate evaluation of algorithm performance in predicting polymorphs. However, we focused only on the MaterialsProject dataset here with specific number of atoms in the unit cell.

Despite outperforming other algorithms by a large margin for regular CSP, it often failed to predict crystals with lower symmetry. However, it successfully predicted space groups of all cubic structures. This is because cubic structures have more symmetry operations and so are easier to predict because symmetry operations reduce the number of distinct configurations that need to be explored. To further improve performance, ParetoCSP2 needs better exploration strategies designed for low-symmetry structures. One possible enhancement is adaptive crossover and mutation strategies, where the algorithm dynamically adjusts crossover and mutation rates based on the complexity of the space group rather than being fixed as a hyperparameter, allowing for more refined searches in high-dimensional configuration spaces. Also, we noticed that though ParetoCSP2 maintained a very high space group diversity, it still declined over generation and ultimately often spanned over mostly higher symmetry space groups. This suggests that the algorithm struggles to preserve specific space group structures due to the influence of genetic operations. To address this, a scheme can be implemented that selects parents from the same space group for mating during genetic operations, which should help preserve a specific space group within the population to sustain high diversity over generations.

ML IAPs such as M3GNet and CHGNet provide notable improvements in prediction speed over DFT, making them a suitable alternative. Although CHGNet is a slightly improved version of M3GNet, it only improved performance over M3GNet in very few cases such as NaGa$_4, $BeSiN$_2$, and PrNi$_2$B$_2$C. However, both of them still face limitations in accurately guiding the optimization process as evidenced by an example in Supplementary Fig. S11 by assigning an intermediate structure a lower energy than the original ground truth structure. For future work, we plan to use improved IAPs, such as eSEN~\cite{esen} and SevenNet~\cite{sevennet} that are trained on more intermediate structures to capture the very small energy differences in the last stages of the optimization process.

\section{Data availability}
The data used in this work can be accessed freely from the MaterialsProject database~\cite{10.1063/1.4812323}.

\section{Code availability}
The source code of this work can be accessed freely at \url{https://github.com/usccolumbia/ParetoCSP2}.

\section{Contribution}
Conceptualization, J.H.; methodology, S.O., J.H. ; software, S.O.; resources, J.H.; writing--original draft preparation, S.O, S.D.,J.H.; writing--review and editing, J.H., S.D.; visualization, S.O.; supervision, J.H.;  funding acquisition, J.H.

\begin{acknowledgement}
The research reported in this work was supported in part by National Science Foundation under the grants 2110033, OAC-2311203, and 2320292. The views, perspectives, and content do not necessarily represent the official views of the NSF.
\end{acknowledgement}

\bibliography{ref}

\end{document}